\documentclass[iop, tighten, numberedappendix, letterpaper]{emulateapj}
 \usepackage{amsmath, amsthm, amssymb}

\newcommand{\hi}{H$\,${\sc i}}
\newcommand{\mh}{$M_{\rm H\textsc{i}}$}
\newcommand{\rh}{${\it R}_{\rm H\textsc{i}}$}

\renewcommand{\thefootnote}{\fnsymbol{footnote}}

\shorttitle{\hi~Power Spectra and the Turbulent ISM of Dwarf Irregular Galaxies}
\shortauthors{Zhang, Hunter \& Elmegreen}

\begin{document}

\title{\hi~Power Spectra and the Turbulent ISM of Dwarf Irregular Galaxies\footnote{
  Based on data from the LITTLE THINGS Survey (Hunter et al., in preparation).}}

\author{
Hong-Xin Zhang\altaffilmark{1,2,3}, Deidre A. Hunter\altaffilmark{2}, Bruce G. Elmegreen\altaffilmark{4}
}
\altaffiltext{1}{Purple Mountain Observatory/Key Laboratory of Radio Astronomy, 
Chinese Academy of Sciences, 2 West Beijing Road, Nanjing 210008, China; hxzhang@pmo.ac.cn}
\altaffiltext{2}{Lowell Observatory, 1400 West Mars Hill Road, Flagstaff, Arizona 86001, USA; hxzhang@lowell.edu; dah@lowell.edu}
\altaffiltext{3}{Graduate School of the Chinese Academy of Sciences, Beijing 100080, China}
\altaffiltext{4}{IBM T. J. Watson Research Center, 1101 Kitchawan Road, Yorktown Heights, New York 10598, USA; bge@us.ibm.com}

\begin{abstract}
\hi~spatial power spectra (PS) were determined for a sample of 24 nearby dwarf 
irregular galaxies selected from the LITTLE THINGS (Local Irregulars That 
Trace Luminosity Extremes -- The \hi~Nearby Galaxy Survey) sample.\ 
The two-dimensional (2D) power spectral indices asymptotically become a constant for each 
galaxy when a significant part of the line profile is integrated.\ For narrow channel maps, the PS 
become shallower as the channel width decreases, and this shallowing trend continues to our single 
channel maps.\ This implies that even the highest  velocity resolution of 1.8 km s$^{-1}$ is not smaller 
than the thermal dispersion of the coolest, widespread \hi~component.\ The one-dimensional PS 
of azimuthal profiles at different radii suggest that the shallower PS for narrower channel width 
is mainly contributed by the inner disks, which indicates that the inner disks have proportionally 
more cooler \hi~than the outer disks.\ Galaxies with lower luminosity ($M_{B}$ $>$ $-$14.5 mag) and star formation 
rate (SFR, log(SFR ($M_\odot\;{\rm yr}^{-1}$)) $<$ $-$2.1) tend to have steeper PS, which implies that the \hi~line-of-sight depths 
can be comparable with the radial length scales in low mass galaxies.\ A lack of a correlation 
between the inertial-range spectral indices and SFR surface density implies that 
either non-stellar power sources are playing a fundamental role in driving the interstellar 
medium (ISM) turbulent structure, or the nonlinear development of turbulent structures has little 
to do with the driving sources.\ 

\end{abstract}


\section{Introduction}
Pervasive interstellar medium (ISM) turbulence regulates star formation (SF) by creating local density 
enhancements and countering gravitational collapse.\ Hierarchical SF (e.g.\ Efremov \& Elmegreen 1998a;  
Gladwin et al.\ 1999; Zhang, Fall \& Whitmore 2001) is a manifestation of a fractal, turbulent 
ISM.\ Kinematic motions from turbulence prevent giant molecular clouds from collapsing 
to stars on the order of a free-fall timescale (Mac Low \& Klessen 2004).\ Turbulent motion is also the dominant 
contributor of the total mid-plane pressure in the solar neighborhood (Boulares \& Cox 1990; 
Jenkins \& Tripp 2001).\ In addition, the stellar initial mass function may be primarily shaped 
by turbulent fragmentation (Padoan \& Nordlund 2002).\ Fast decay ($\sim$ a crossing time across 
the driving scale, Stone et al.\ 1998; Mac Low 1999) of the turbulent energy suggests 
a continuous driving mechanism is in operation.\ It has been suggested that among the various power 
sources for turbulence, such as magnetorotational instabilities (MRI, e.g.\ Sellwood \& Balbus 1999), 
gravitational instabilities (e.g.\ Wada et al.\ 2002), thermal instabilities (e.g.\ Kritsuk \& Norman 2002), 
and stellar energy (e.g.\ winds, supernovae), supernovae dominate the energy input to the ISM (e.g.\ 
Norman \& Ferrara 1996; Mac Low \& Klessen 2004).\ Nevertheless, as pointed out by Elmegreen \& Scalo (2004), 
gravitational energy, which has an energy input rate an order of magnitude lower than that of supernovae, 
may have a higher efficiency for conversion into turbulence.\ Indeed, recent hydrodynamic galaxy 
simulations (Bournaud et al.\ 2010; Hopkins et al.\ 2012) found that ISM turbulence can be driven 
by gravitational processes on scales larger than or comparable to the Jeans length, with stellar 
feedback on smaller scales being essential also in maintaining a stable ISM cloud structure.\ 
Observationally, it remains to be seen how ISM turbulence is related to SF.\ 

Fourier transform power spectra, which characterize the relative importance of structures 
at different scales, have been extensively used as a diagnostic for ISM structures.\ A power-law 
behavior of the power spectra of \hi-emission line intensities was found in the Milky Way 
(e.g.\ Crovisier \& Dickey 1983; Green 1993; Dickey et al.\ 2001; Khalil et al.\ 2006), 
Small Magellanic Cloud (SMC, Stanimirovi\'c et al.\ 2000), and Large Magellanic Cloud 
(LMC, Elmegreen et al.\ 2001).\ The power-law power spectrum is usually attributed to 
ISM turbulence.\ In unmagnetized, incompressible Kolmogorov turbulence (Kolmogorov 1941), 
kinetic energy is injected at large scale (the driving scale), and significant dissipation 
by viscosity occurs only at small scales (dissipation scale).\ The scale range in between 
the driving scale and dissipation scale, which has a power-law energy spectrum ($E(k)$ 
$\sim$ $k^{-5/3}$), is known as the inertial range.\ The kinetic energy injected on large 
scales cascades to small scales without much loss in the inertial range.\ 

In incompressible, subsonic turbulence, density fluctuations passively follow the velocity field, 
which obeys Kolmogorov scaling (Lithwick \& Goldreich 2001).\ In other words, density and 
velocity have similar Kolmogorov power spectra $P \sim$ $k^{-11/3}$, where $k$ is the three-dimensional 
(3D) wavevector.\ However, the transonic (e.g.\ warm neutral medium, WNM) or supersonic 
(e.g.\ cold neutral medium, CNM) nature of the ISM implies compressible turbulence.\ The 
ISM turbulence also involves self-gravity and magnetic fields.\ Numerical simulations (Kim \& Ryu 2005; 
Kowal, Lazarian, \& Beresnyak 2007; Gazol \& Kim 2010) suggest that the spatial power spectrum 
is sensitive to the sonic Mach number and the Alfv\'en Mach number, in the sense that higher 
sonic Mach number and Alfv\'en Mach number (i.e.\ weaker magnetic forces) lead to steeper velocity 
power spectra and shallower density power spectra due to the formation of stronger shocks.\ 
Krumholz \& McKee (2005) proposed that the sonic Mach number $\mathit{M}$ is a  
factor in determining the star formation rate (SFR $\propto$ $\mathit{M}^{-0.32}$).
 
\hi~is an important component of the ISM.\  
Especially in gas-rich dwarf irregular (dIrr) galaxies, neutral \hi~usually dominates the baryonic component 
(e.g.\ Zhang et al.\ 2012).\ Observationally, the intensity fluctuations of individual \hi~channel maps are caused 
by both 3D real space projection and velocity mapping.\ Lazarian \& Pogosyan (2000, hereafter LP00) found 
that the intensity fluctuations within thin velocity slices (less than the turbulent velocity dispersion at the studied 
scale) of the observed position-position-velocity (PPV) data cubes are generated or significantly affected by 
the turbulent velocity field, whereas intensity fluctuations in thick velocity slices are dominantly caused by 
density fluctuations because the velocity fluctuations are averaged out.\ Stanimirovi\'c \& Lazarian (2001) 
applied this velocity channel analysis technique to the \hi-emission line data of the SMC.\ They found that the 
power-law power spectra become steeper with increasing velocity slice width, from which the power 
spectral indices were derived for both the density and velocity fields.\  

To investigate the relationship between the spatial power spectra and SF, we present a velocity channel 
analysis for a subsample of LITTLE THINGS dIrr galaxies.\ The paper is structured as follows.\ 
In Section 2 we briefly describe the \hi-emission line data used in this work.\ The power spectrum variations with 
channel width are presented in Section 3.\ Section 4 gives a comparison between our 
image-domain-based and the visibility-based estimation of power spectra in the literature.\ In Section 5, 
we give the indication of the trend that power spectra become shallower as the channel width gets narrower.\ 
Section 6 explores various correlations between power spectral indices and SF-related quantities.\ 
Discussion about obtaining velocity spectral indices from our observations is given in Section 7.\
The summary follows in Section 8.\

\section{Sample and Data}
The 24 dIrr galaxies studied in this work (Table 1) were selected from the LITTLE THINGS sample 
(Hunter et al., in preparation) which were drawn from the large optical survey of dwarf 
galaxies by Hunter \& Elmegreen (2006).\  The LITTLE THINGS is a large NRAO Very 
Large Array (VLA\footnote[6]{The VLA is a facility of the National Radio Astronomy 
Observatory (NRAO), itself a facility of the National Science Foundation operated under 
cooperative agreement by Associated Universities, Inc.}) project, which was granted nearly 
376 hours of VLA time in the B-, C- and D-array configurations to perform 21-cm \hi-emission 
line observations of a representative sample of 41 nearby ($D \lesssim$ 10 Mpc) dIrr galaxies.\ 
The sub-sample of 24 galaxies is chosen to be relatively face-on (inclination $<$ 55$^{\circ}$), 
and these galaxies cover a large range of galactic parameters, such as integrated luminosity 
($-$18 $<$ $M_{B}$ $<$ $-$10 mag), central surface brightness (18.5 $<$ $\mu_{0}^{V}$ 
$<$ 25.5 mag arcsec$^{-2}$), SFR surface density ($-$4 $<$ 
log($\Sigma_{\rm SFR} (M_\odot\;{\rm yr}^{-1}\;{\rm kpc}^{-2})$) $<$ $-$1.3), and atomic gas 
richness (0.2 $<$ $M_{\rm gas}/M_{*}$ $<$ 26, see Zhang et al.\ 2012).\ Note that all but 
three (DDO 50, NGC 3738, NGC 4214, Table 1) of our galaxies are fainter than the SMC 
($M_{B}$ = $-$16.35 mag, Bekki \& Stanimirovi\'c 2009).

To better handle extended emission, LITTLE THINGS adopted the multi-scale {\sc clean} 
({\sc msclean}) algorithm implemented in the Astronomical Image Processing System (AIPS).\ 
In particular, four different scale sizes, i.e.\  0$''$, 15$''$, 45$''$ and 135$''$, were chosen for 
mapping both small- and large-scale emission (Hunter et al., in preparation).\ For this work, 
we use the \hi~data cubes in the image-domain created with the {\sc robust} = 0.5 weighting scheme in the AIPS 
task {\sc imagr}.\ The cubes are cleaned down to a flux level of 2 times the rms noise, determined 
in line-free channels.\ After the cleaning and mapping processes, we use the task {\sc blank} to 
remove the noise.\ Briefly speaking, we first convolve the cube to a spatial resolution of 25$''$, 
then in this low-resolution cube only regions containing emission ($>$ 2 -- 3 $\sigma$) in at 
least three consecutive channels are considered as areas of real emission and the rest is blanked 
(set to zero), and then we apply this Master Blanking cube to blank our full-resolution cube.\

The typical spatial resolution of the {\sc robust} maps is $\sim$ 7$''$.\ Thirteen galaxies have 
a channel width of 1.3 km s$^{-1}$, with a real velocity resolution of 1.8 km s$^{-1}$.\ The other 11 
galaxies have a channel width of 2.6 km s$^{-1}$, with a velocity resolution of 2.6 km s$^{-1}$.\ 
Limited by the shortest baseline of the VLA's compact array configuration, our observations are 
blind to emission from structures with angular scales $>$ $\sim$ 15$'$.\ For our sample, except 
for IC 10, all the galaxies have an \hi~extent (the largest scale for power spectrum analysis) well 
below 15$'$.\ The VLA primary beam is 32$'$ at 21 cm.\ The data cubes in the image-domain were 
corrected for primary beam attenuation, and then de-projected using geometrical parameters 
(galactic center, position angle and axial ratio) derived from ellipse fitting to the velocity-integrated 
\hi~maps.\ An intrinsic axial ratio of 0.3 (Hodge \& Hitchcock 1966) was adopted when converting 
the axial ratios to inclination angles.\ We point out that adopting a slightly different intrinsic axial 
ratio introduces negligible uncertainties to our results in this work.\

\section{Power spectrum variations with channel width}

\begin{figure*}[htb!]
\begin{center}
\includegraphics[width=0.9\textwidth]{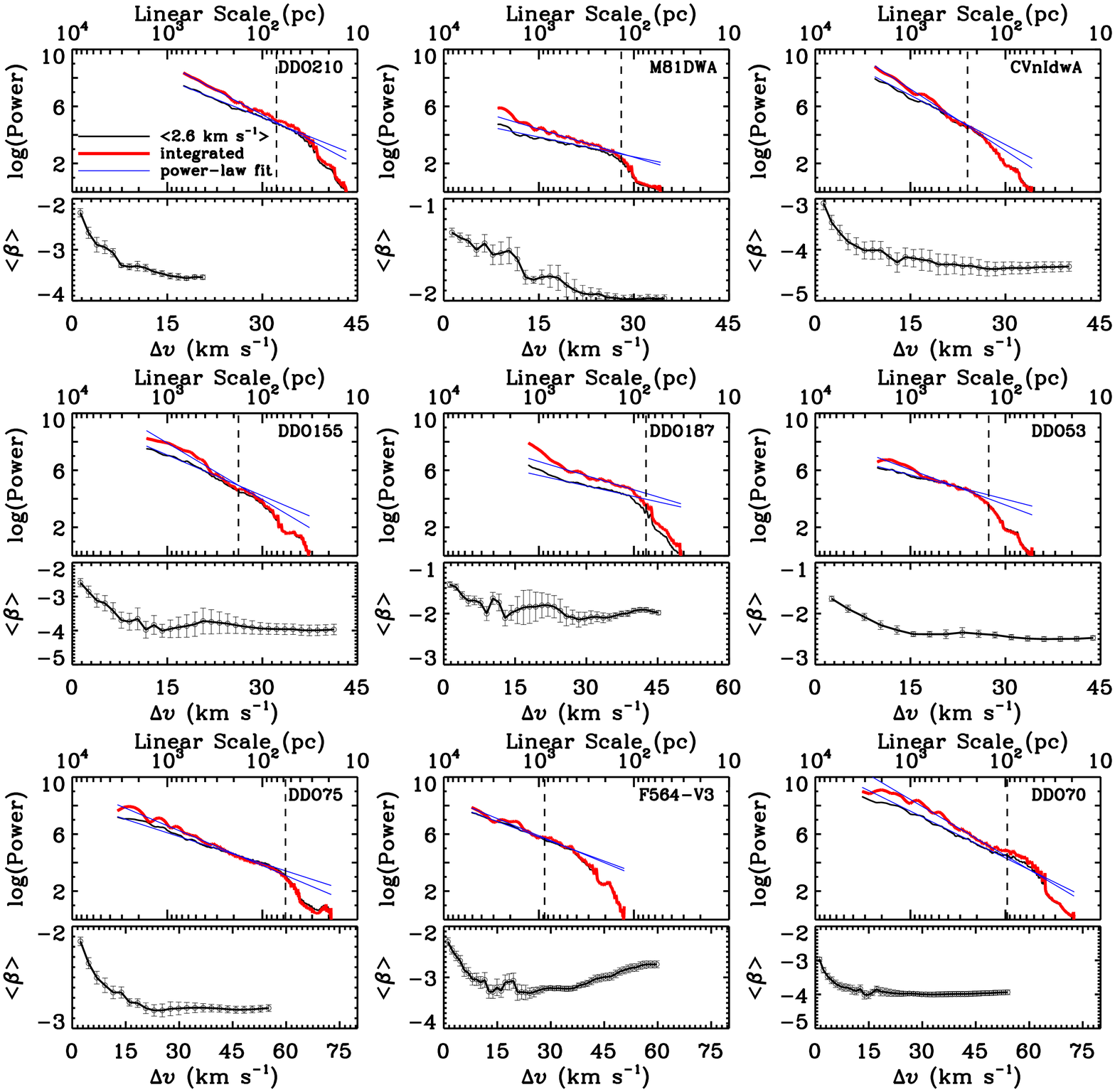}
\caption{Power spectrum variations as a function of channel width.\ The average power spectra 
of 2.6 km s$^{-1}$ thick channel maps ({\it thin black solid line}), and power spectra of velocity-integrated 
maps ({\it thick red solid line}) are shown in the upper panel of each galaxy plot.\ The power-law fit to the 
power spectrum is overplotted as {\it thin blue solid line}.\ The power spectra have been arbitrarily shifted 
vertically.\ The vertical {\it dashed} lines mark the linear scales of the synthesized beam.\ Note that finite 
size of the synthesized beam causes the decline of power on linear scales smaller than the 
beam size.\ Variations of power-law spectral indices with channel width (in velocity units) are shown 
in the lower panel of each galaxy plot.
\label{fig1}}
\end{center}
\end{figure*}

We determined the two-dimensional (2D) spatial power spectra of 
velocity slices with different width, by gradually rebinning the channels 
with bin size increasing from the single channel width of 1.3 km s$^{-1}$ 
(or 2.6 km s$^{-1}$) to a sum over whole line profiles.\ Line-emission channels 
with at least 20 per cent of the peak flux of the global \hi~velocity profile were 
used in the analysis.\ The power spectra were obtained using the Fast Fourier 
Transform algorithm.\ Note that the power spectra presented in this paper are 
the average within each annulus (1 spatial frequency unit wide) in the 2D wavenumber 
$k$ ($k_{x}, k_{y}$) space.\ The finite size of the synthesized beam results in a 
pronounced decline at high spatial frequencies of the power spectrum (Figure \ref{fig1}).\ 
Also, toward the lowest spatial frequencies of the power spectrum, some galaxies exhibit 
an obvious up-bending trend compared to higher spatial frequencies, which may be caused 
by some large-scale symmetric structures in the galaxies.\ Therefore, we fit each power 
spectrum with a power law ($P \propto k^{\beta}$) for linear scales from 1.5 times the 
beam size up to the point where the power spectrum starts bending upward.\ 

For a given slice width, the power spectral indices of individual velocity slices with 
the same width are averaged to get the average spectral index $<\beta>$ for that width.\ 
Most of our galaxies have non-zero inclinations, which means that an  
individual channel map shows only part of the galaxy, due to the galactic rotation.\ 
Thus galactic rotation may restrict the lowest spatial frequencies of the power spectra 
of narrow channel maps.\ If the fitted power spectral indices with and without the lowest 
one or two spatial frequencies are significantly different, we removed the lowest one 
or two spatial frequencies from the power spectrum fitting.\ As an example, Figure \ref{fig2} 
shows some 2.6 km s$^{-1}$ channel maps ({\it black contours}) of DDO 133 overlaid on 
the integrated map (grey scale image).\ Our result suggests that, at a given channel width, 
different channel maps have about the same power spectrum, although they may show 
different parts of the galaxy.\ 

\begin{figure*}[htb!]
\begin{center}
\includegraphics[width=0.9\textwidth]{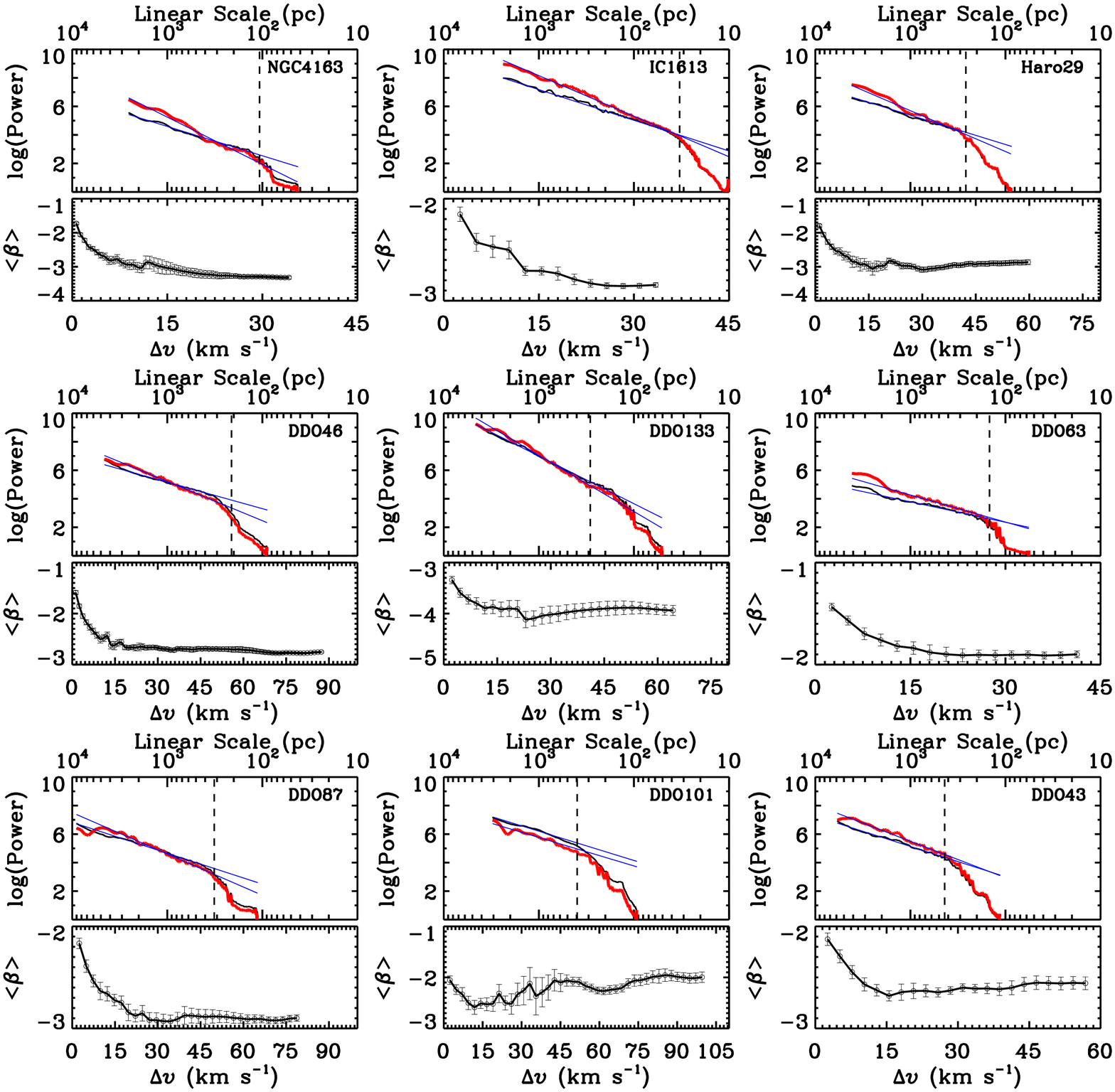}
Fig.~1--{\it Continued}
\end{center}
\end{figure*}

\begin{figure*}[htb!]
\begin{center}
\includegraphics[width=0.9\textwidth]{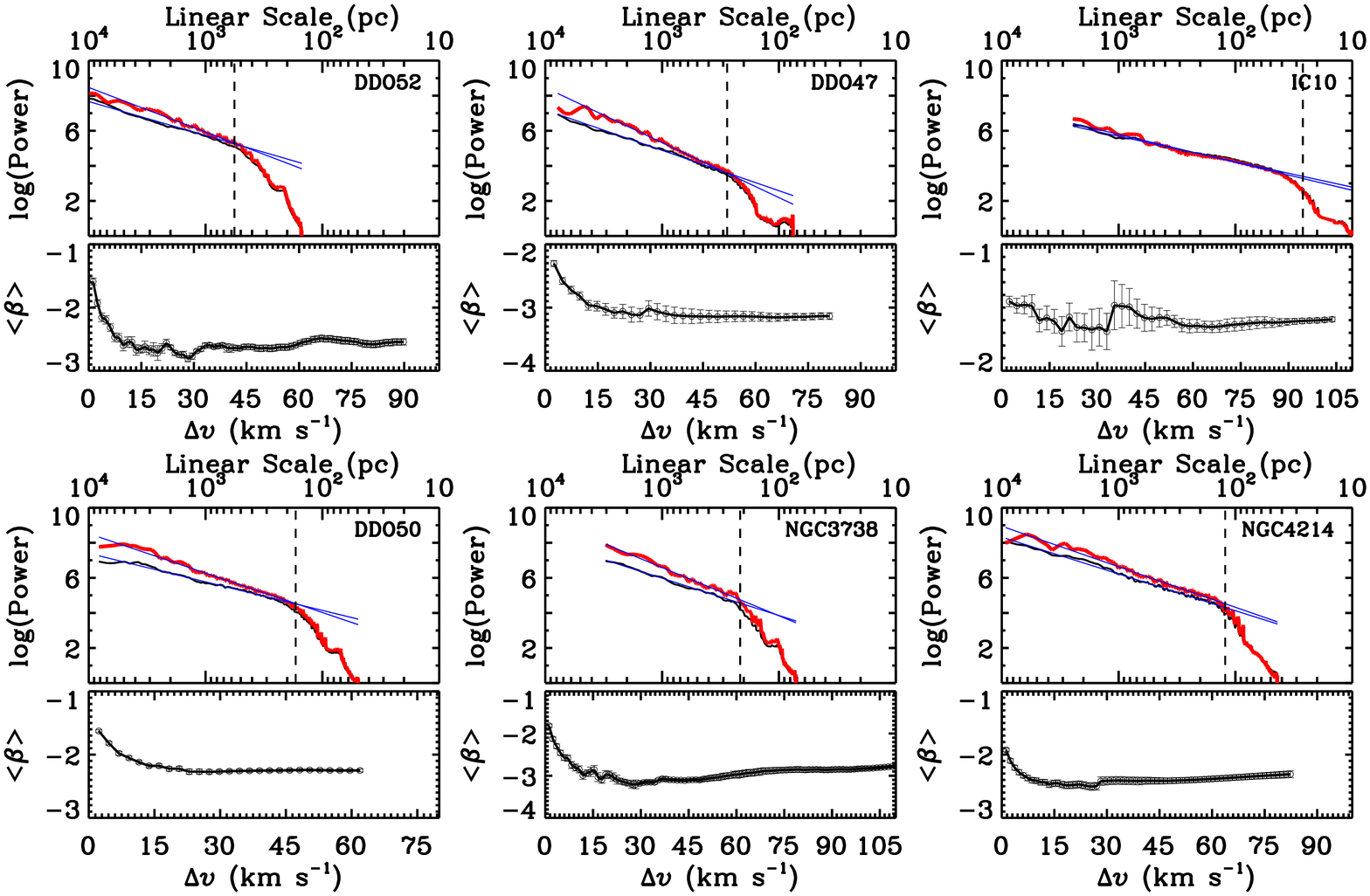}
Fig.~1--{\it Continued}
\end{center}
\end{figure*}

The variations of $<\beta>$ with increasing slice width 
are shown in Figure \ref{fig1}.\ The error bars in Figure \ref{fig1} are determined as 
the standard deviation of $\beta$ divided by $\sqrt{n}$, where $n$ is the number of 
slices that were averaged together.\ In Figure \ref{fig1}, we also present the average 
power spectra of the 2.6 km s$^{-1}$ channel maps ({\it thin black solid lines}) and the 
power spectra of the velocity-integrated maps ({\it thick red solid lines}).\ The {\it vertical 
dashed lines} in Figure \ref{fig1} mark the linear scales of the 
synthesized beam.\ Table 2 lists the linear ranges used in the power-law fitting, the 
spectral indices of the velocity-integrated and the 1.3 km s$^{-1}$ and 2.6 km s$^{-1}$ 
channel maps, the approximate slice width at which the power spectra start getting 
shallower for smaller slice widths, and the velocity spectral indices, discussed below.\   

\begin{figure*}[htb!]
\begin{center}
\includegraphics[width=0.9\textwidth]{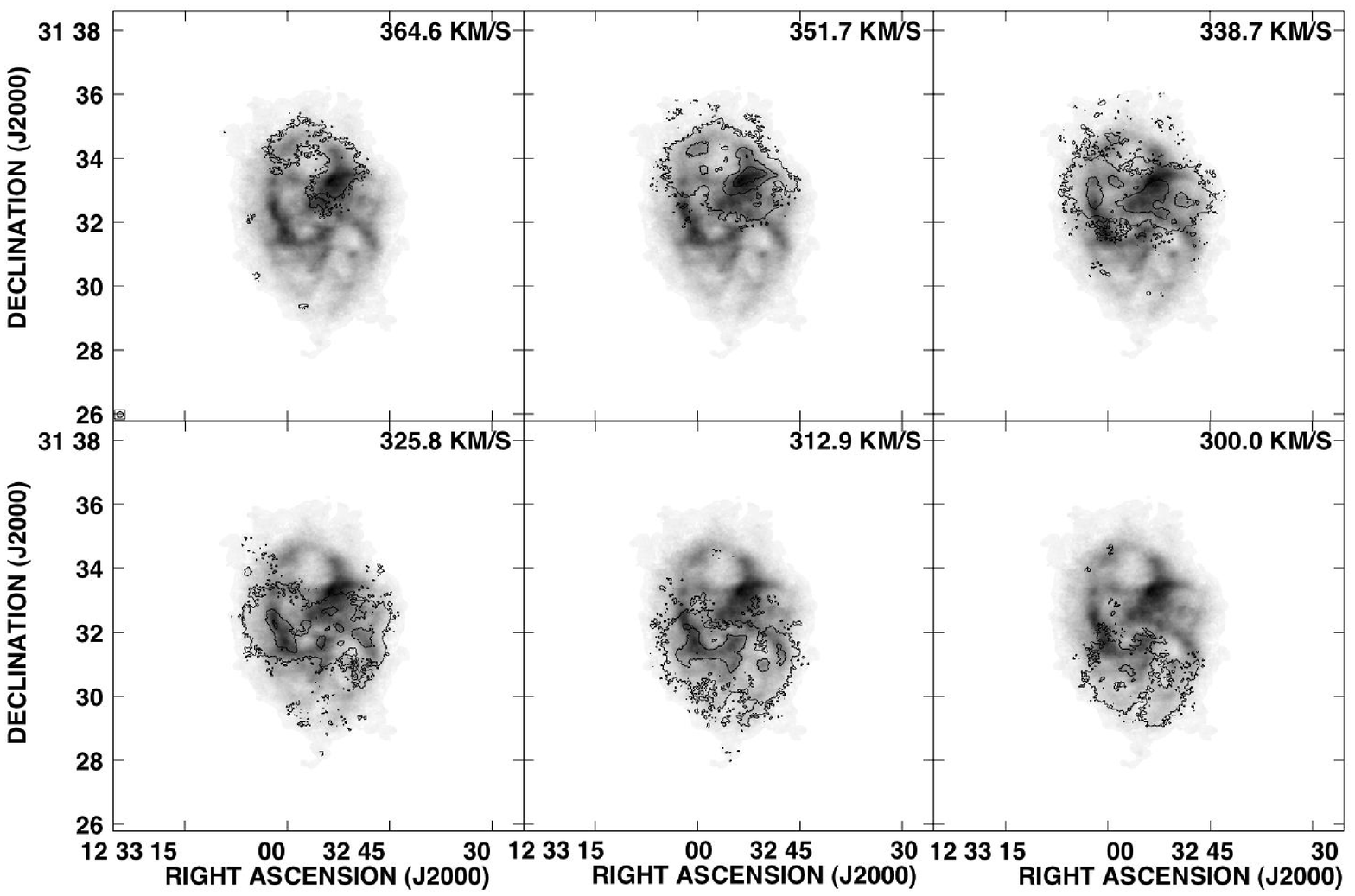}
\caption{
Every fifth 2.6 km s$^{-1}$ channel map of DDO 133 used in our power spectrum 
analysis is shown as {\it black contours}.\ The {\it grey scale} image in each panel 
is the integrated intensity map.\ The synthesized beam is indicated at the left-bottom 
corner of the upper-left panel.\ The contour levels are 2, 10, and 20 times the rms noise.\
\label{fig2}}
\end{center}
\end{figure*}

As shown in Figure \ref{fig1}, the power spectra are described very well by a single power law 
on large scales.\ Except for the two nearest galaxies (i.e.\ IC 1613, IC 10), the robust 
linear scales of the derived power spectra are $\gtrsim$ 100 pc.\ DDO 187 and F564-V3 have the 
smallest range (less than a factor of 5) of linear scales used in the power spectrum fitting.\ As the slice 
width decreases, the power spectra of F564-V3 and DDO 101 get steeper first, then get shallower.\  
Except for F564-V3 and DDO 101, the others exhibit gradually steeper power spectra as the velocity 
slice gets thicker, and the spectral index approaches a constant as a significant part of the line 
profile is integrated together.\ 

The turbulent velocity field is correlated with spatial scales, in the sense that the velocity dispersion   
on larger scales is usually larger than that on smaller scales.\ In incompressible turbulence, the 
turbulent density field passively follows the velocity field.\ In a narrow channel map, small-scale 
structures from independent gas elements on the line of sight combine to give relatively 
more power on smaller scales.\ This results in a shallower power spectrum for a narrower channel map.\ 
However, if the spatially-uncorrelated thermal velocity is non-negligible in a given channel map, 
then small-scale structures with turbulent velocity dispersions smaller than the thermal speed 
will lose their distinction.\ If the thermal velocity is much larger than the turbulent velocity dispersion 
even on the largest scales, then the thermal velocity component will totally smooth out the spatial 
correlation present in the turbulent velocity field.\ Therefore, the shallowing trend for power spectra 
of narrower channel maps is a reflection of the spatially-correlated turbulent velocity field, modulated 
by the thermal velocity.\  

According to the theoretical calculations of LP00, the power-law indices of turbulent density and velocity 
power spectra can be obtained by changing the width of velocity slices of the PPV data cube.\ If the effective 
velocity slice width, which is jointly determined by the instrumental channel width and the thermal 
velocity (see below), is significantly smaller than the turbulent velocity dispersion on the studied scales, 
then the velocity slice is regarded to be thin, and the power spectral indices remain constant as 
the slice width further decreases.\ If the effective velocity slice width is significantly larger than the 
turbulent velocity dispersion on the studied scales, then the velocity slice is regarded to be thick, and the 
power spectral indices remain constant as the slice width further increases.\ The spectral indices of 
thick slices are determined solely by the turbulent density fluctuations.\ In the $\beta$-$\Delta\upsilon$ plot 
of Figure \ref{fig1}, the asymptote on the side of thick velocity slices suggests that density fluctuations 
determine the intensity fluctuations there.\ The shallower power spectra of intensity fluctuations in 
narrower velocity slices implies more influence from velocity fluctuations.\

\section{Comparison with a visibility-based estimation of power spectra}
Dutta et al.\ (2009b) have determined \hi~power spectra of a sample of dwarf galaxies using a different method.\
They used data from the Giant Metrewave Radio Telescope (GMRT), and the range of 
baselines roughly corresponds to those of the VLA D and C array configurations.\ 
They determined the power spectra directly from the observed visibilities in the $uv$ plane
rather than producing maps first and then taking the Fourier Transform and determining the power spectra.\
This avoids an unnecessary step (converting visibilities into maps) since the 
$uv$ data are already in the Fourier domain.\ Here we prefer to identify real emission 
from the target by working in the image plane.\ Also, the correction for inclination and primary 
beam can only be properly done in the image plane.\ 

The Dutta et al.\ (2009b) study has two galaxies in common with our sample: DDO 210 
and DDO 155.\ They found that DDO 210 has power spectral indices of  $-$2.1 $\pm$ 0.6 
and $-$2.3 $\pm$ 0.6 for velocity-integrated and single channel (1.7 km s$^{-1}$) velocity 
slices, respectively, and DDO 155 has power spectral indices of $-$0.7 $\pm$ 0.3 and 
$-$1.1 $\pm$ 0.4 for velocity-integrated and single channel velocity slices, respectively.\ 
These are compared to our values of $-$3.66 $\pm$ 0.05 and $-$2.17 $\pm$ 0.09 for DDO 210, 
and $-$3.97 $\pm$ 0.15 and $-$2.59 $\pm$ 0.12 for DDO 155.\ Recall that the single channel 
of 1.3 km s$^{-1}$ is narrower than the real velocity resolution of $\sim$ 1.8 km s$^{-1}$.\  
The power spectra of DDO 155 are not power-law for linear scales above 
1.5 times the beam size, and thus the quality of power-law fitting is very poor (Figure \ref{fig1}).\ 
We notice that the power spectra of DDO 155 determined by Dutta et al.\ (2009b) show 
similar behavior on large scales.\ So DDO 155 may not be a good case for comparison.\ 
For DDO 210, our single channel spectral index is consistent with that determined by 
Dutta et al.\ within the uncertainty of their measurement.\ Our velocity-integrated 
spectral index of DDO 210 is much steeper than that determined by Dutta et al.\ (2009b).\
Based on figure 8 of Dutta et al.\ (2009b), the multi-channel-integrated power spectrum 
of DDO 210 is a very good power law for linear scales above $\sim$ 0.18 kpc, 
below which the spectrum becomes very noisy and (thus) flat.\ Dutta et al.\ included the 
noisy, flat part of the power spectrum in their power-law fitting, which leads to a 
much less negative spectral index.\

The key difference between the image-domain-based and visibility-based power spectra  
lies in the different ways used to avoid noise bias.\ Noise present in the visibilities or 
images shallows the derived power spectrum.\ To reduce the effect of noise in the $uv$ plane, 
Dutta et al.\ determined the power spectrum by correlating the visibilities at slightly 
different baselines for which the noise is assumed to be uncorrelated.\ This method 
works on the assumption that the angular extent of the galaxy is much smaller 
than the primary beam of the telescope.\ The compromise present in this method is that, on 
the one hand, sufficient visibility pairs are needed to get good statistics and on the other hand, the 
baseline differences should be as small as possible ($<1/\theta_{0}$, where $\theta_{0}$ is the 
angular extent of the target galaxy) to have strong enough correlations between visibilities of different 
baselines.\  

In the image plane, we distinguish between real emission and noise primarily in two steps.\
First, in the dirty map, only peaks 2 $\sigma$ above the noise level are regarded as signal and 
{\sc msclean}-ed.\ Second, as described above, only regions with flux $>$ 2 -- 3 $\sigma$ in 
at least three consecutive channels in the smoothed cubes are regarded as real emission, and 
the rest is blanked.\ We found that the power spectral indices of both DDO 210 and DDO 155 
would be about $-$1.5 regardless of the channel width if we do not apply blanking to the cubes.\
This implies that non-blanked cubes are predominantly affected by noise on small scales.\ 
{\sc msclean} has been shown to work excellently in recovering extended structures 
(Cornwell 2008; Rich et al.\ 2008).\ Compared to the classical {\sc clean}, {\sc msclean}, 
with its scale-sensitive nature, can clean down to the noise level without leaving skewed 
noise on the residual map.\ Furthermore, the recovered fluxes from {\sc msclean} (and {\sc blank}) were 
found to be consistent with those from single-dish observations (Hunter et al., in preparation).\ 
Nevertheless, the noise present in the $uv$ data can definitely affect the non-linear deconvolution 
involved in the image restoration process.\

Given these different approximations and assumptions in the two methods, it is 
not obvious to tell which method is better in determining power spectra.\ 
In this work, however, we point out that, 1) by including the deep VLA B-array data, 
the outer part (longer baseline length) of the $uv$ plane is better sampled in our data; 
2) the two galaxies in common all have non-negligible inclinations.

\section{Effective velocity slice width and multi-phase neutral \hi}
The distinction between the thin and thick regimes depends on a comparison 
between the squared turbulent velocity dispersion on the studied scale $\sigma_{turb,~l}^{2}$ 
and the squared effective velocity slice width $\delta V^{2}$ (LP00) 
\begin{equation}
\begin{split}
\sigma_{turb,~l}^{2} \ll  \delta V^{2},~{\rm thick} \\ \sigma_{turb,~l}^{2} \gg  \delta V^{2},~{\rm thin}
\end{split}
\end{equation}
The thermal motion ($v_{T}$), which is spatially incoherent, acts like a smoothing along the 
velocity dimension.\ Thermal broadening, together with the instrumental 
channel width, determine the effective velocity slice width.\ Assuming a uniform sensitivity 
across individual channels, the effective velocity slice width is given by 
\begin{equation}\label{eq1}{\delta V \sim 2(\Delta\upsilon^{2}/6+2v_{T}^{2})^{1/2}}\end{equation}
where $\Delta\upsilon$ and $v_{T}$ are the individual channel width and typical thermal velocity, 
respectively (LP00).\ It is the effective velocity slice width, rather than the channel width, that 
determines the effective slice thickness, and thus the variation of power spectral indices.\ 

The thermal velocity of the gas restricts the minimum effective velocity slice 
width that can be achieved.\ For an isothermal gas with temperature $T$, 
the effective velocity slice width remains nearly constant when the channel width 
becomes smaller than the thermal velocity $v_{T}$ ($\sqrt{2k_{\rm B}T/m\pi}$), 
and thus the power spectral indices do not change.\ Empirically, the median channel 
width at the point where the power spectra start to get shallower for narrower channel 
width is $\sim$ 15 km s$^{-1}$ for our sample galaxies (Table 2).\ A two-phase 
(WNM and CNM) description of the neutral \hi~was suggested by Field, Goldsmith, 
\& Habing (1969).\ Wolfire et al.\ (2003) demonstrate that the interstellar \hi~at a 
temperature of $\sim$ 100 K and of $\sim$ 8000 K can coexist in thermal equilibrium.\ 
Therefore, if there is only thermally stable WNM, the power spectral indices would 
not change for channel widths narrower than $\sim$ 6.5  km s$^{-1}$ (the 
thermal velocity of \hi~gas with a temperature of 8000 K).\ For the thirteen galaxies 
with a channel width of 1.3 km s$^{-1}$, the power spectra keep getting shallower 
for narrower channel width, down to 1.3 km s$^{-1}$.\ This suggests widespread 
\hi~with a temperature $\sim$ 600 K (corresponds to a thermal velocity of 1.8 km s$^{-1}$) 
or even lower, considering that the real velocity resolution is 1.8 km s$^{-1}$.\ In the Milky 
Way, Heiles \& Troland (2003) found that more than 48\% of the WNM, which accounts for 
$\sim$ 60\% of the total \hi, is in the thermally unstable regime with a broad temperature 
range from $\sim$ 500 -- 5000 K.\ A temperature of  $\sim$ 600 K lies in the broad boundary 
between the WNM and CNM.\  

The typical velocity dispersion from \hi~emission line observations of nearby 
star-forming galaxies is $\sim$ 10 km s$^{-1}$ (e.g.\ C\^ot\'e, Carignan \& Freeman 2000; 
Leroy et al.\ 2008), which is the combination of turbulent and thermal velocity dispersions.\ 
Therefore, $\sim $ 10 km s$^{-1}$ can be regarded as an upper limit on the turbulent 
velocity on galactic scales.\ Also, since the maximum temperature $T_{\rm max}$ of 
neutral \hi~is $\sim$ 10$^{4}$ K, the minimum possible turbulent velocity dispersion 
$\sigma_{turb}$ should be $\sim$ 4 km s$^{-1}$ ($\sqrt{10^{2}-v_{T_{\rm max}}^{2}}$).\
For a given scale, the thin (thick) regime is reached if the squared effective slice width is 
much smaller (larger) than the squared turbulent velocity dispersion (Inequality 1).\ 
In the $\beta$ vs.\ $\Delta\upsilon$ plot, two asymptotes are expected 
for the two regimes of thin and thick effective velocity slices.\ As mentioned above, 
the \hi~gas with a temperature of $\sim$ 600 K has a thermal velocity $v_{T}$ $\sim$ 
1.8 km s$^{-1}$ which is the highest spectral resolution for our data.\ With $v_{T}$ = 1.8 
km s$^{-1}$ and $\Delta\upsilon$ = 1.8 km s$^{-1}$, Equation 2 gives an effective velocity 
slice width of $\sim$ 5 km s$^{-1}$.\ Since we do not see the asymptotic behavior towards 
the single channel width in the $\beta$ vs.\ $\Delta\upsilon$ plot (Figure \ref{fig1}), the 
single channel, and thus the effective velocity slice width of 5 km s$^{-1}$ 
does not reach the thin regime, at lease for the thermally unstable WNM components 
with a temperature $\lesssim$ 600 K.\ Based on Inequality 1, the turbulent velocity 
dispersion $\sigma_{turb,~l}$ for the neutral \hi~gas with a temperature $\lesssim$ 600 K 
is not much higher (i.e.\ by less than a factor of 3) than $\sim$ 5 km s$^{-1}$, which is mildly 
supersonic. 

The line profiles of our galaxies cover a velocity range from $\sim$ 20 to 130 km s$^{-1}$.\ 
The velocity-integrated slices lie in the thick slice regime, as implied by the asymptotes 
seen in Figure \ref{fig1}.\ As mentioned above, our galaxies start reaching the thick regime 
at a channel width of $\sim$ 15 km s$^{-1}$ on average.\ With $v_{T}$ = 6.5 km s$^{-1}$ 
($T_{\rm WNM, stable}$ = 8000 K) and $\Delta\upsilon$ = 15 km s$^{-1}$, Equation 2 gives 
an effective velocity slice width of 22 km s$^{-1}$.\ Therefore, to be in the thick regime defined 
in Inequality 1, the turbulent velocity dispersion of the thermally stable WNM should be much 
smaller (i.e.\ by at least a factor of 3) than 22 km s$^{-1}$.\ Therefore, the maximum Mach number 
for the thermally stable WNM components should be smaller than $\sim$ 1 
($\sigma_{turb,~l}$/$v_{T_{\rm WNM, stable}}$, $\sigma_{turb,~l}$ $<$ 22/3).\ 

The intensity fluctuations of the velocity-integrated 
maps are determined by density fluctuations including all possible phases of \hi.
On the other hand, as explained above, the shallowing trend for power spectra of 
channel maps narrower than 6.5 km s$^{-1}$ is only caused by turbulent velocity 
fluctuations of cooler \hi, including the thermally unstable components and (possibly) 
some CNM.\ The large \hi~self-absorption survey by Gibson et al.\ (2005) found a smoothly 
distributed, albeit fluffy, cold phase of \hi~in the Milky Way Galaxy, which is in line with our finding 
that \hi~gas with a temperature of 600 K or even lower is widespread in dIrr galaxies.

\begin{figure*}[htb!]
\begin{center}
\includegraphics[width=0.9\textwidth]{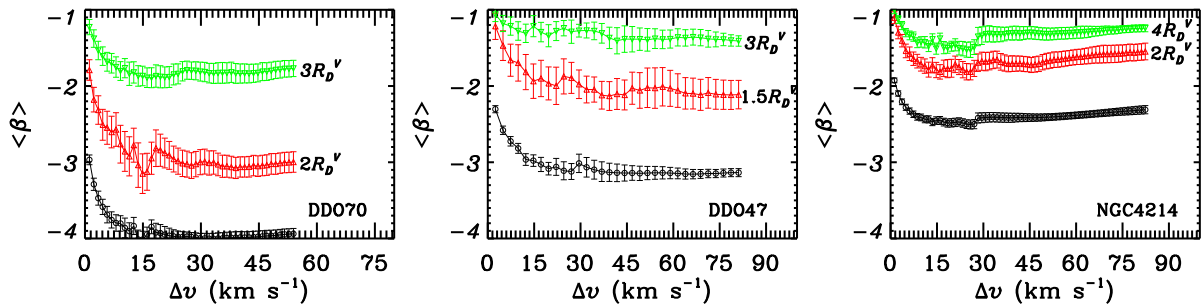}
\caption{Variations of 1D spectral indices as a function of channel width for three galaxies.\ 
The {\it red} triangles denote the spectral indices of azimuthal scans at an inner radius, measured 
in units of $V$-band disk scalelength.\ The {\it green} upside down triangles denote the spectral 
indices of azimuthal scans at an outer radius.\ The 2D spectral indices presented in Figure \ref{fig1} 
are shown as {\it black} circles.\
\label{fig3}}
\end{center}
\end{figure*}

Zhang et al.\ (2012) found that star-forming disks of most local dIrr galaxies have been 
shrinking at least during the past $\sim$ Gyr.\ Cold atomic gas is the precursor to molecular 
cloud formation which shows an almost linear correlation with the SFR in nearby star-forming 
galaxies (e.g.\ Leroy et al.\ 2008).\ Possible radial variations of the $\beta$-$\Delta\upsilon$ 
relation would reflect the relative spatial distributions of  \hi~gas with different temperatures.\ 
We took one-dimensional (1D) power spectra of azimuthal scans at different radii (e.g.\ 
Elmegreen et al.\ 2003) for our galaxies.\ Figure \ref{fig3} presents the results for three of our 
galaxies (DDO 70, DDO 47 and NGC 4214).\ The 1D power spectra in Figure \ref{fig3} are 
the average of adjacent azimuthal scans with a radial extent of two synthesized beams.\ 
We used the same geometric parameters derived above for the extraction of azimuthal 
profiles.\ The {\it black} symbols correspond to the 2D power spectra as shown in Figure \ref{fig1}.\ 
The {\it red} and {\it green} symbols, respectively, denote the spectral indices at inner and 
outer radii, in units of the $V$-band disk scalelength $R_{D}^{V}$.\ In an isotropic field, the 
1D power spectra of azimuthal profiles are expected to be shallower than the 2D power 
spectra by $\sim$ 1 because of the reduced dimension (see the power spectra of inner 
radii in Figure \ref{fig3}).\ The azimuthal scans at the outer radii have systematically 
shallower power spectra than those at the inner radii, which may signify a gradual change 
from 3D to 2D geometry, possibly due to the longer azimuthal scans at the outer radii.\ 
Obviously, the inner regions exhibit a much stronger shallowing trend toward narrower 
channel width.\ This suggests that the shallowing trend observed for the 2D power spectra, 
and thus the structure associated with the cooler \hi, is dominantly contributed by the inner, 
more actively star-forming regions.\ The more widespread cool \hi~in the inner disk may 
be caused by a higher mid-plane gas pressure (Wong \& Blitz 2002; Blitz \& Rosolowsky 2004) 
or a higher average gas volume density (Gao \& Solomon 2004) there.\ Braun (1997) studied the 
neutral \hi~properties of a sample of 11 nearby spiral galaxies.\ He found positive radial gradients 
of the \hi~kinematic temperature, and that there are more \hi~components with narrow 
($\lesssim$ 6 km s$^{-1}$) emission line profiles in the inner disks.

\section{Power spectral index vs. SF} 
The ISM structure in dIrr galaxies can be significantly influenced by stellar feedback.\ 
Shells and holes of up to $\sim$ kpc scales are often seen in the \hi~gas distribution of 
dwarf galaxies (e.g.\ Sargent et al.\ 1983; Puche et al.\ 1992; Young \& Lo 1997; 
Walter \& Brinks 1999; Ott et al.\ 2001; Muller et al.\ 2003; Simpson, Hunter \& Knezek 2005; 
Cannon et al.\ 2011).\ It has been shown that stellar feedback can provide sufficient energy 
to produce the observed shells and holes (Warren et al.\ 2011).\  Stellar energy, such as 
winds, ionizing radiation, and supernova explosions, dominates the energy injection into the 
ISM (Mac Low \& Klessen 2004) on the corresponding scales, at least in star-forming galaxies.\ 
By studying the visibility-based power spectra of seven dwarf galaxies, Dutta et al.\ (2009b) found 
that the galaxies with higher SFR surface density tend to have steeper \hi~power spectra, implying 
more large-scale structures of \hi~for more intense SF.\ With a much larger sample, we can explore 
the possible relationship between the power spectra and SF.\

For most of the galaxies (Table 2), the minimum linear scales covered by our 
power spectra are $\gtrsim$ 100 pc, which is probably comparable with the expected disk 
scale height of dIrr galaxies (e.g.\ van den Bergh 1988; Staveley-Smith et al.\ 1992; 
Carignan \& Purton 1998; Elmegreen, Kim \& Staveley-Smith 2001; Banerjee et al.\ 2011).\ 
A break in the power-law power spectrum of the \hi~emission distribution has been 
observed for a few galaxies (LMC -- Elmegreen et al.\ 2001; NGC 1058 -- Dutta et al.\ 2009a).\ 
The power spectral index for smaller scales (compared to the break scale) is steeper than for 
larger scales by $\sim$ 1, which is found both in the LMC (Elmegreen et al.\ 2001) and simulations 
of gas-rich galaxies (Bournaud et al.\ 2010).\ The flattening of the power spectrum on 
larger scales is understood as the transition from isotropic 3D turbulence on smaller 
scales to anisotropic 2D turbulence on larger scales (Bournaud et al.\ 2010).\ The break is 
about the gas disk scale height (Elmegreen et al.\ 2001; Bournaud et al.\ 2010).\ 

\begin{figure}[htb!]
\begin{center}
\includegraphics[width=0.5\textwidth]{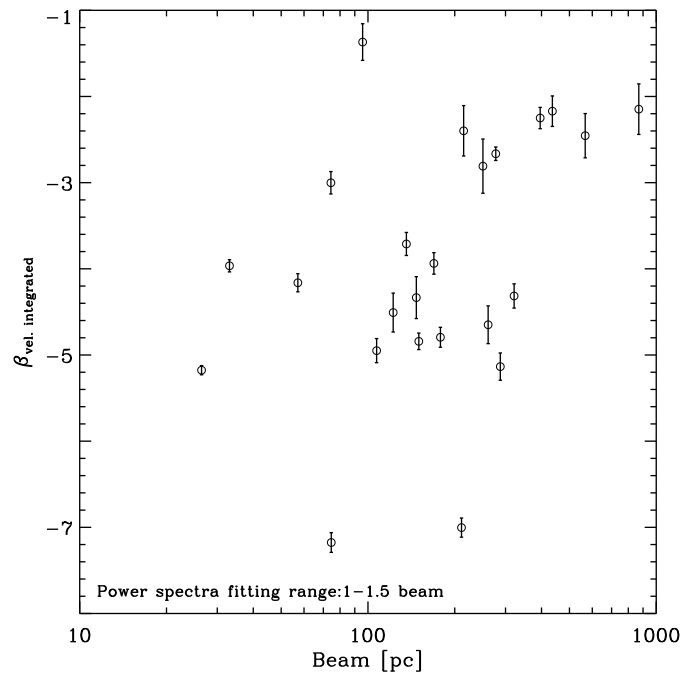}
\caption{
The synthesized beam size (pc) is plotted against the velocity-integrated power spectral indices 
determined by fitting a linear scale range from 1 to 1.5 times the beam size.\ 
\label{fig4}}
\end{center}
\end{figure}

As described in Section 3, we did power law fitting to the power spectra for linear 
scales larger than 1.5 times the synthesized beam.\ This is because the finite size 
of the synthesized beam results in a pronounced decline at high spatial frequencies 
of the power spectra (Figure \ref{fig1}).\ Nevertheless, we expect the power spectra at 
high spatial frequencies, although being noticeably influenced by the beam, still partially 
reflect the intrinsic power spectra of the galaxy at the resolution limit.\ Figure \ref{fig4} 
presents the relation between the beam size (in units of pc) and the velocity-integrated 
power spectra fitted over linear scales from 1 to 1.5 times the beam size.\ Our galaxies 
fall into two groups in Figure \ref{fig4}, separated around a beam size of $\sim$ 200 -- 300 
pc, below which the galaxies have more negative indices.\ The two groups are also roughly 
divided by a spectral index of $-$3.\ This may reflect the effect of the disk thickness, because 
anisotropic 2D turbulence on large scales is expected to have much shallower power 
spectra than isotropic 3D turbulence on small scales.\

The SMC, which is like the most luminous galaxies in our sample in terms of stellar 
and gas masses, does not have a break in the power spectrum between linear scales 
of 30 pc and 4 kpc (Stanimirovi\'c et al.\ 1999).\ This lack of a break indicates that 
the maximum transverse scale is comparable to or smaller than the \hi~line-of-sight depth.\ 
Perhaps the 3D behavior of the power spectrum of the SMC across a large range of linear scales is 
a result of the tidal interaction with the LMC during the past few Gyr.\ The numerical simulations by 
Bekki \& Chiba (2007) found that almost 20\% of the SMC's gas may have been accreted by the 
LMC during their recent interaction, and this may explain the origin of the LMC's intermediate-age 
stellar populations with distinctively low metallicities (e.g.\ Geisler et al.\ 2003).\ So it is possible that 
the SMC's gas disk may have been stretched during the recent interaction, resulting in a thicker disk.\ 
For the two nearest galaxies in our sample, IC 10 and IC 1613, the power spectra also 
can be well fitted with a single power law between linear scales of $\sim$ 50 pc and $\sim$ 2 kpc.\ 
IC 1613 has a tidal index $\sim$ 0.9 (Karachentsev et al.\ 2004), which indicates that this galaxy 
has been significantly influenced by the neighboring galaxies (e.g.\ M 33).\ 
Our \hi~kinematics analysis (Oh et al.\ in preparation) suggests that these two galaxies, and also 
the very low luminosity galaxies, such as DDO 210 and DDO 155, have ratios of the maximum 
rotational velocity to velocity dispersion $\lesssim$ 2, which indicates that the gas distribution 
may be a relatively thick disk or even triaxial ellipsoid.\ We note that, Roychowdhury et al.\ (2010, 
see also S\'anchez-Janssen et al.\ 2010) found a mean intrinsic axial ratio of $\sim$ 0.6 for 
the \hi~disks of dIrr galaxies with $M_{B}$ $>$ $-$14.5 mag, which suggests very thick gas 
disks in faint dIrr galaxies.\ 

\begin{figure*}[htb!]
\begin{center}
\includegraphics[width=0.9\textwidth]{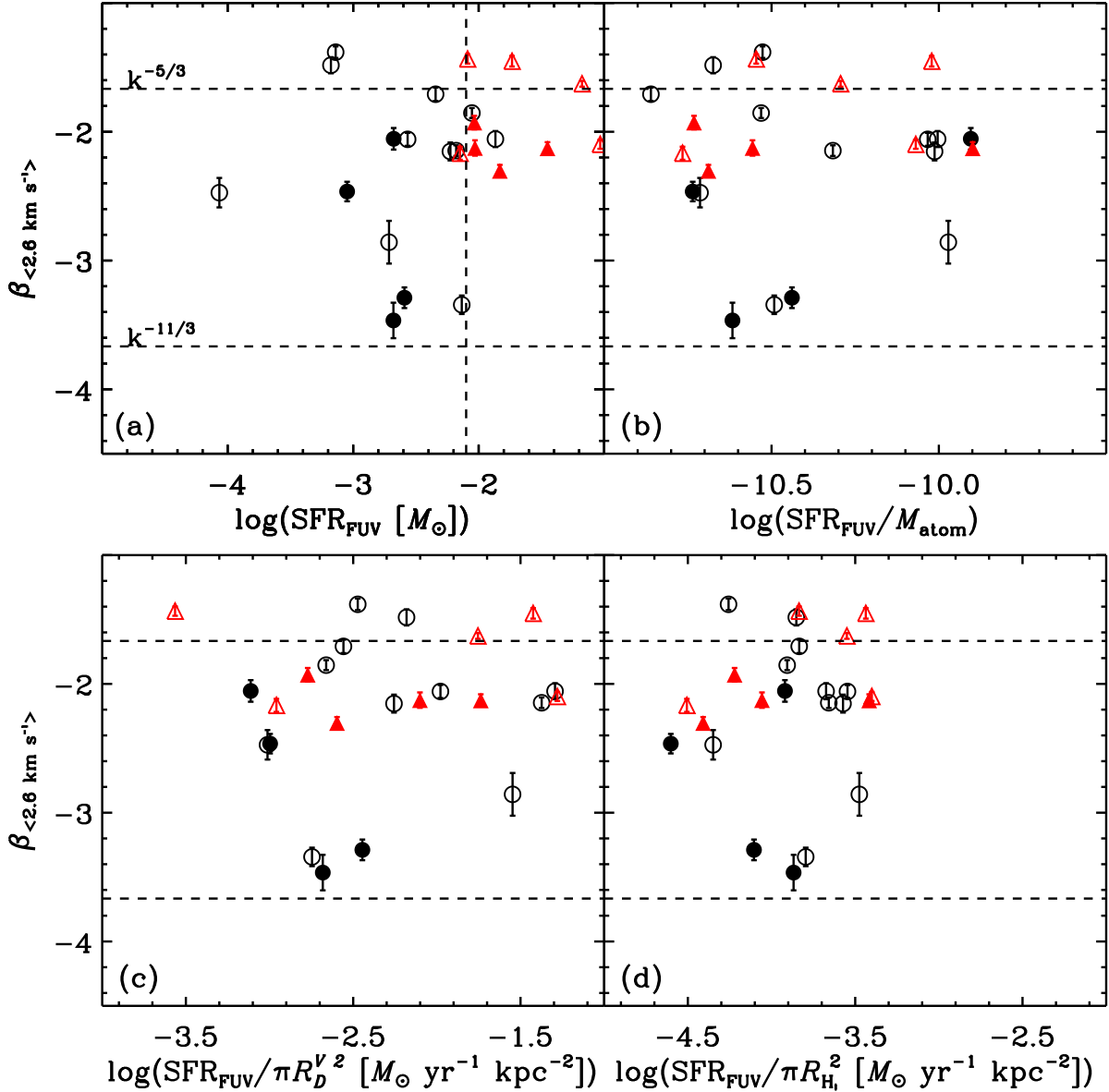}
\caption{Power spectral indices of 2.6 km s$^{-1}$ channel maps are plotted against the 
SFR determined from the FUV (a), SF efficiency (b), SFR normalized by the area enclosed 
by the radius of one $V$-band disk scale length (c), and SFR normalized by the area enclosed 
by the radius where the \hi~column density falls off to 10$^{19}$ cm$^{-2}$ (d).\ 
The {\it open} and {\it filled} symbols respectively denote the galaxies with spectral indices 
measured at high spatial frequencies (linear scales between 1 and 1.5 times the beam size) 
more negative and less negative than $-$3.\ The {\it black circles} and {\it red triangles} respectively 
denote the galaxies with $M_{B}$ $>$ $-$14.5 mag and $M_{B}$ $<$ $-$14.5 mag.\ 
The horizontal dashed lines mark the expected 1D ($P \sim k^{-5/3}$) and 3D ($P \sim k^{-11/3}$) 
power spectral indices of a Kolmogorov-type turbulence.\ The vertical dashed line in panel (a) 
marks the division (log(SFR$_{\rm FUV}$) = $-$2.1) of the bimodality discussed in the text.\ 
We notice that there is a slight trend for the galaxies with $M_{B}$ $<$ $-$14.5 mag in panel (d), in the 
sense that higher SFR surface density corresponds to shallower power spectra, 
but the scatter is too large to be sure. 
\label{fig5}}
\end{center}
\end{figure*}

\begin{figure*}[htb!]
\begin{center}
\includegraphics[width=0.9\textwidth]{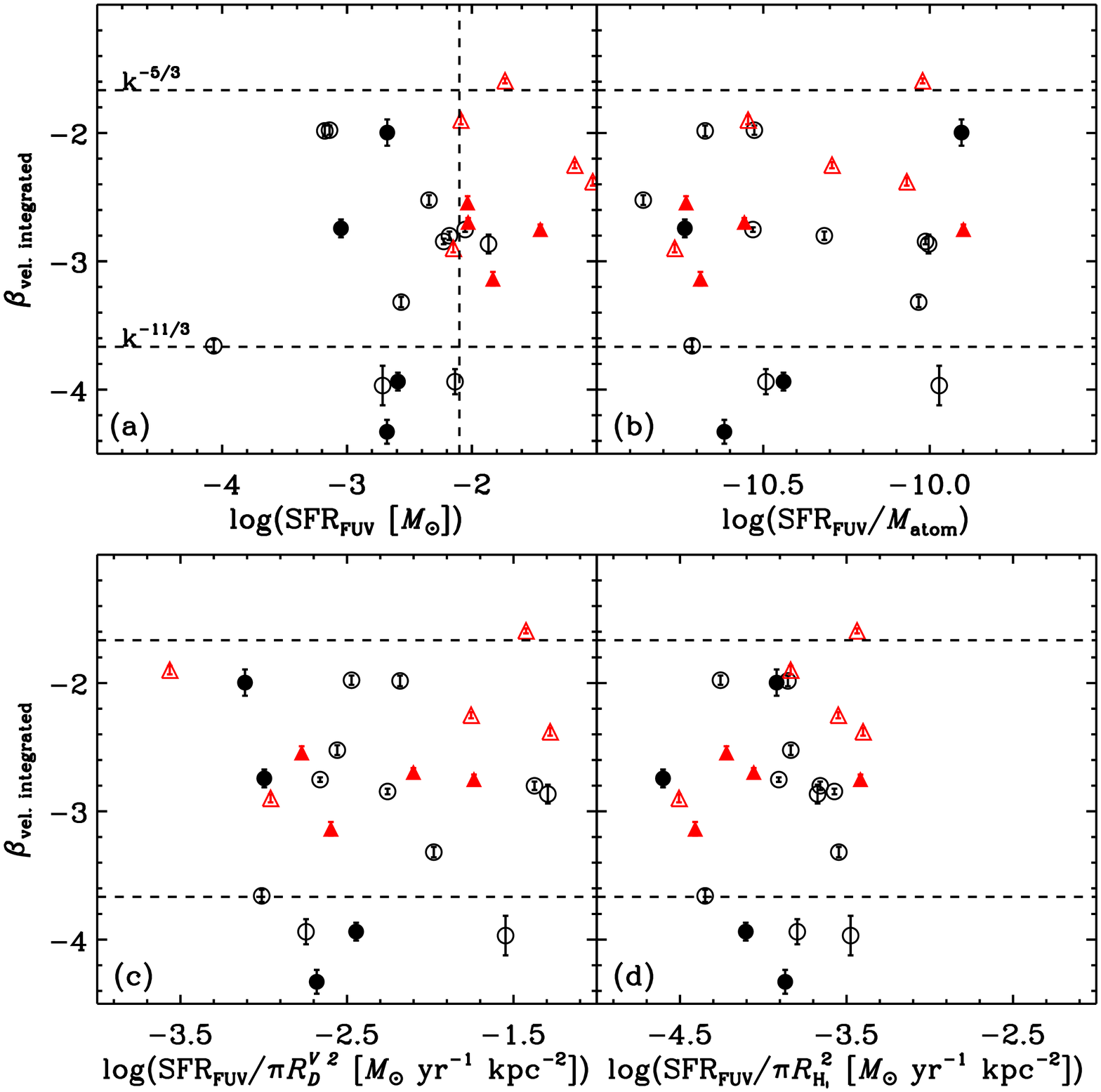}
\caption{
Same as Figure \ref{fig5} but for velocity-integrated maps.\ 
The slight correlation found in Figure 5d is also present here in panel (d).  
\label{fig6}}
\end{center}
\end{figure*}

Figures \ref{fig5} and \ref{fig6} plot the power spectral indices against SF-related quantities derived from the 
far-UV (FUV, SFR$_{\rm FUV}$): SF efficiency of the atomic gas (SFR$_{\rm FUV}$/$M_{\rm atom}$), 
SFR surface density normalized to the area within one $V$-band disk scalelength (SFR/$\pi{{R_{D}^{V}}^{2}}$), 
and SFR surface density within the radius where the \hi~column density falls off to $10^{19}$ 
cm$^{-2}$(SFR/$\pi$\rh$^{2}$), for 1.3 and 2.6 km s$^{-1}$ thick slices and for the 
velocity-integrated maps, respectively.\ For the formula used to derive SFR$_{\rm FUV}$ from the FUV, the 
reader is referred to Hunter et al.\ (2010) and Zhang et al.\ (2012).\ The total atomic gas mass 
$M_{\rm atom}$ (1.34$\times$ \mh~to account for He) used here was collected from single-dish 
observations in the literature (see Hunter \& Elmegreen 2004 for the references).\ 
The {\it open} and {\it filled} symbols respectively denote the galaxies with spectral indices 
measured at high spatial frequencies (1--1.5 times the beam) more negative and less 
negative than $-$3, i.e.\ the two groups mentioned above.\ The {\it black circles} and 
{\it red triangles} respectively denote the galaxies with $M_{B}$ $>$ $-$14.5 mag and 
$M_{B}$ $<$ $-$14.5 mag.\ There is no strong correlation in these plots.\ However, 
Figure \ref{fig5} suggests that all the galaxies with log(SFR) $>$ $-$2.1 have narrow 
channel spectral indices around $-$2, whereas galaxies with lower SFR exhibit a 
much larger scatter toward more negative spectral indices.\ The {\it vertical dashed} 
line in panel (a) of Figure \ref{fig5} marks the division at log(SFR) = $-$2.1.\ Similarly, 
in Figure \ref{fig6}, the spectral indices of velocity-integrated maps exhibit a larger 
scatter (toward more negative indices) in galaxies with lower SFR.\ 

\begin{figure*}[htb!]
\begin{center}
\includegraphics[width=0.9\textwidth]{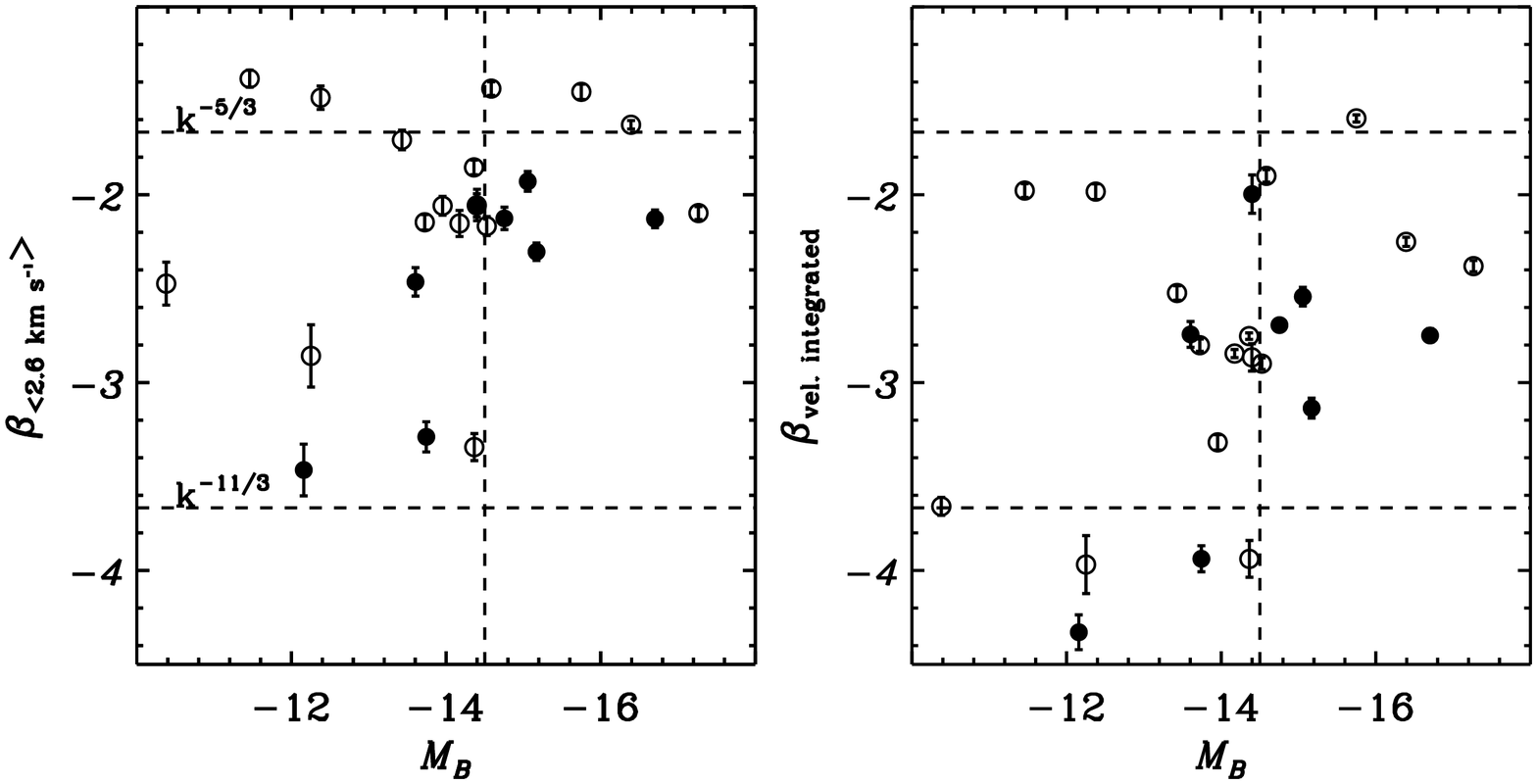}
\caption{Power spectral indices of 2.6 km s$^{-1}$ channel maps ($left~panel$) and 
velocity-integrated maps ($right~panel$) are plotted against $B$-band absolute 
magnitude $M_{B}$.\ The {\it open} and {\it filled} symbols respectively denote 
the galaxies with spectral indices measured at high spatial frequencies (linear scales 
between 1 and 1.5 times the beam size) more negative and less negative than $-$3.\ 
The horizontal dashed lines mark the expected 1D ($P \sim k^{-5/3}$) and 3D ($P \sim k^{-11/3}$) 
power spectral indices of a Kolmogorov-type turbulence.\ The vertical dashed lines 
mark the division of the bimodality ($M_{B}$ = $-$14.5 mag) discussed in the text. 
\label{fig7}}
\end{center}
\end{figure*}

Given that the spectral indices show the bimodality with the SFR but not the SFR surface density, 
the bimodality may just reflect the correlation between spectral indices and the mass 
(or size) of galaxies.\ In Figure \ref{fig7}, the spectral indices of 2.6 km s$^{-1}$ channel maps 
and spectral indices of velocity-integrated maps are plotted against $B$-band absolute 
magnitude $M_{B}$.\ There exists a similar bimodality as that present between power 
spectral indices and the SFR.\ The {\it vertical dashed} lines in Figure \ref{fig7} mark the 
division of the bimodality at $M_{B}$ = $-$14.5 mag.\ Brighter galaxies tend to have flatter 
power spectra.\ This could be a result of disk thickness, in the sense that brighter galaxies 
tend to have a higher ratio of radial length to thickness, and so are more 2D on average.\ 
The three galaxies which have the most negative spectral indices (for both narrow channel 
maps and velocity-integrated maps) are CVnIdwA, DDO 70 and DDO 133.\ The negative 
tidal indices of these three galaxies (Karachentsev et al.\ 2004) suggest that they do not 
have significant interactions with neighboring galaxies.\  

This ratio of radial length to thickness also affects the regularity of galactic structure because 
the thickness is approximately the turbulent Jeans length, and therefore the typical size 
for stellar complexes.\ When the radial length is large compared to the thickness, there is 
room in the disk for a lot of star-forming clumps, giving the disk a somewhat uniform 
appearance.\ But when the radial length is small compared to the thickness, there may 
be only a few star-forming complexes with large stochasticity, which can lead to 
an irregular appearance of the disk.\ Thus the correlation we find between the slope 
of the power spectrum and the galaxy absolute magnitude (or total SFR) may explain 
the findings of Lee et al.\ (2007), who explored the distribution of the Local Volume 
galaxies in the $M_{B}$-H$\alpha$ equivalent width plane.\ They found a bimodality around 
$M_{B} \sim $ $-$15 mag in the sense that galaxies with lower luminosity exhibit a larger (by a factor 
of 2) scatter in H$\alpha$ equivalent width than galaxies with higher luminosity 
($-$19 $ \lesssim M_{B} \lesssim $ $-$15 mag).

The lack of a strong correlation between the inertial range spectral indices and SFR surface 
density (and also SF efficiency) may be unsurprising.\ In the classical Kolmogorov turbulence, 
the energy transfer occurs locally (cascade) in the wavenumber space in the inertial range, 
and the driving only affects the energy input at the top of the cascade.\ Unlike this ideal 
Kolmogorov turbulence which has a single driving source on the large scale,  
the driving of turbulence in the ISM may span a wide range of scales.\ The stellar energy can 
drive turbulence from parsec-scale supernovae up to hundred-parsec-scale 
superbubbles (Mac Low \& Klessen 2004).\ Similarly, gravitational energy may feed the turbulence 
through sub-parsec gravitational collapse, up to galactic-scale gravitational instabilities 
(e.g.\ Wada et al.\ 2002; Elmegreen et al.\ 2003; Agertz et al.\ 2009) and galaxy interactions 
(e.g.\ Elmegreen \& Scalo 2004), etc.\ Therefore, the single power-law behavior of \hi~power spectra 
suggests that the turbulence may be driven (by whatever energy sources) over a wide range 
of physical scales, otherwise the power spectra should exhibit features, such as a flattening at low 
wavenumbers, at the primary driving scale (e.g.\ Nakamura \& Li 2007; Padoan et al.\ 2009).\ 
Supernovae are the largest contributers of energy input to the ISM on scales comparable to or 
smaller than the disk thickness (Mac Low \& Klessen 2004).\ However, it was suggested that 
supernova driven turbulence alone cannot explain the broad \hi~emission lines, at least in the 
outer parts of disk galaxies (e.g.\ Dib, Bell \& Burkert 2006; Tamburro et al.\ 2009).\ Furthermore, 
simulations (e.g.\ Agertz et al.\ 2009; Bournaud et al.\ 2010) found that gravitational instabilities 
alone can reproduce the observed power-law power spectra, the stellar feedback does not 
significantly change the ISM statistical properties established by gravitational instabilities.\ 
In addition, the multi-phase nature of the ISM implies that thermal instabilities (e.g.\ Field 1965) 
may also be an important driving agent of turbulence (e.g.\ Hennebelle \& Audit 2007; 
Gazol \& Kim 2010).\ Thus turbulence may be driven by many sources and the properties 
of this turbulence on scales smaller than the sources may be independent of these 
sources, so a correlation between spectral index and SFR surface density might not 
be expected.

The slope of the power spectrum may depend on the nature of the driving force 
(no matter whether it is stellar or non-stellar).\ Recent simulations done by Federrath, Klessen, 
\& Schmidt (2009) suggest that, at a given Mach number, compressive forcing can lead to 
steeper density spectra than solenoidal forcing (rotational, incompressible).\  
Kim \& Ryu (2005), Kowal, Lazarian, \& Beresnyak (2007), and Gazol \& Kim (2010) found a 
correlation between power spectral index and Mach number from magnetohydrodynamic (MHD) 
simulations, in the sense that higher sonic Mach number leads to shallower density power 
spectra.\ This correlation was interpreted as the result of more small-scale density structures 
generated by stronger shocks in supersonic flows.\ In the present observations, the lack of 
a correlation between the power spectral indices and the SFR surface density may imply that 
stellar feedback and other energy sources share similar characteristics of driving on average.\  
Burkhart et al.\ (2010) also found little correlation between SF activity and Mach number in a 
2D map based on kurtosis of \hi~line profiles in the SMC.\ They found that regions with the 
highest sonic Mach number lie around the bar.\ Chappell \& Scalo (2001) arrived at a similar 
conclusion using a multifractal spectrum analysis of low-mass star-forming cloud complexes: 
there is little correlation between the geometrical properties of the gas and the level of internal SF. 

As we discussed above, higher-luminosity galaxies ($M_{B}$ $<$ $-$14.5 mag) in our sample may 
have 2D turbulence on the studied scales and 3D turbulence on unresolved scales, whereas 
lower-luminosity galaxies may have only 3D turbulence because the disks are thick relative to 
their radial length scales.\ Therefore, it may be desirable to explore the 
relationship between SFR surface density and power spectral indices for 
higher-luminosity and lower-luminosity galaxies separately.\ Figures 5 and 6 
suggest that our claim of a lack of a correlation between SFR surface density 
and spectral indices is valid for galaxies with both $M_{B}$ $<$ $-$14.5 mag or 
$M_{B}$ $>$ $-$14.5 mag.\ The lack of a correlation for the more luminous galaxies 
makes sense according to the above discussion if their power spectra are dominated 
by 2D turbulence and stellar feedback has little effect on scales larger than the disk 
thickness.\ The lack of a correlation for less luminous galaxies, which have relatively 
thicker disks than more luminous galaxies, suggests that local SF does not 
strongly affect the spectral index of turbulence.\

\section{The turbulent velocity field}
Unlike the spectra that reflect density fluctuations, velocity spectra (specific kinetic 
energy spectra) are directly related to the turbulent energy distribution across different 
scales.\ Intensity fluctuations within channel maps are contributed by both the density 
and velocity fields.\ However, the relevant importance of density and velocity fields 
changes with the amplitude of the density fluctuations.\ For example, intensity fluctuations 
simply follow the velocity field if the density field is constant.\ Our thick velocity slices, 
especially the velocity-integrated ones, have intensity fluctuations dominated by variations 
in the density field.\ According to LP00, the spectral index of (2D) thick velocity slices  
$\beta_{\rm thick}$ (e.g.\ $\beta_{\rm integrated}$ listed in Table 2) is equal to the 3D 
density spectral index $\beta_{\rm density}$, provided the maximum transverse scale 
is comparable to or smaller than the line-of-sight depth.\ In an isotropic 3D turbulent 
medium, if the density spectral index $< -3$, the velocity spectral index $\beta_{\rm velocity}$ 
is equal to $-2\beta_{\rm thin}-9$, where $\beta_{\rm thin}$ is the spectral index of intensity 
fluctuations within thin velocity slices.\ $\beta_{\rm velocity}$ is equal to 
$-3-2(\beta_{\rm thin} - \beta_{\rm thick})$ if the density spectral index $> -3$.

As discussed in Section 4, the velocity resolution of 2.6 km s$^{-1}$ does not yet reach 
the thin slice regime.\ Assuming the 2.6 km s$^{-1}$ velocity slice is thin 
(i.e.\ $\beta_{\rm thin}$ = $\beta_{\rm 2.6}$) and the line-of-sight depth is comparable 
to the maximum transverse scale we studied, we can obtain the velocity spectral indices 
$\beta_{\rm velocity}$ which are listed in Table 2.\ Sixteen galaxies have velocity power 
spectra steeper than that of Kolmogorov turbulence (for which $\beta_{\rm velocity}$ would 
be $-$11/3 for 3D turbulence).\ The real velocity spectral indices can be even steeper 
because we do not yet reach the ``thin'' regime.\ If the line-of-sight depth is smaller than 
the maximum transverse scale, the measured $\beta_{\rm thick}$ and $\beta_{\rm thin}$ 
should be subtracted by $\sim$ 1 to be used for the calculation of $\beta_{\rm density}$ and 
$\beta_{\rm velocity}$, and thus the resultant $\beta_{\rm density}$ would be $\beta_{\rm thick}$ 
$-$ 1, and $\beta_{\rm velocity}$ would be shallower than those listed in Table 2 by $\sim$ 2 when 
the $\beta_{\rm density}$ $<$ $-$3, whereas $\beta_{\rm velocity}$ 
determined for the galaxies with $\beta_{\rm density} > -$ 3 is not affected (see the relations in the 
preceding paragraph).\ We point out that, if the steepening factor of 2D turbulence relative 
to 3D turbulence is not 1, then we would not know how to obtain $\beta_{\rm density}$, and, 
in the case of $\beta_{\rm thick}$ $<$ $-$3, $\beta_{\rm velocity}$ when the line-of-sight 
depth is smaller than the maximum transverse scales.\  

MHD isothermal simulations (e.g.\ Kritsuk et al.\ 2007; Federrath et al.\ 2009; Gazol \& Kim 2010) 
suggest that fluids with higher Mach number (thus stronger compressibility) have steeper velocity 
power spectra (and shallower density power spectra).\ However, the isothermal assumption adopted in most 
simulations may not be valid for the multi-phase \hi~for which thermal instabilities can be an important 
driving mechanism of turbulence (Hennebelle \& Audit 2007; Gazol \& Kim 2010).\ We emphasize that 
the multi-phase nature of the \hi~makes the velocity spectral indices derived here questionable, 
because the density power spectra determined from thick velocity slices may invoke density fields of all different 
phases of \hi, whereas the turbulent velocity fields as reflected by the shallower power spectra within narrower 
velocity slices only include contributions from the thermally unstable WNM and (possibly) some CNM.

\section{Summary}
We have studied the \hi~power spectral index variations with channel width for 
a sample of nearby dIrr galaxies.\ The majority of the 2D power spectra cover more than one 
decade of linear scales, from $\sim$ hundred pc to several kpc.\ 
The main results are summarized as follows.\
\begin{enumerate}

\item 
The power spectral indices asymptotically become a constant for each galaxy when a significant 
part of the line profile is integrated, consistent with the theoretical calculations of LP00.\ 
This indicates that density fluctuations, including all possible temperature 
components of \hi, determine the intensity fluctuations of our ``thick'' velocity slices.
\item
Starting at a channel width of $\sim$ 15 km s$^{-1}$ on average for our sample, narrower 
channel maps have shallower power spectra.\ The shallowing trend, which is caused in part by turbulent 
velocity dispersions of the thermally unstable WNM and possibly some CNM, continues down 
to the single channel maps (1.3 km s$^{-1}$).\ This continuation indicates that, first, even the 
highest velocity resolution of 1.8 km s$^{-1}$ is not smaller than the thermal dispersion of the 
coolest \hi~($\lesssim$ 600 K) which is widespread in our galaxies; if it were, then the spectral 
index would remain constant.\ Second, the turbulent velocity dispersion of the coolest \hi~($\lesssim$ 600 K) 
probed at our highest velocity resolution is not much larger than $\sim$ 5 km s$^{-1}$, which 
means that the turbulence in this phase of \hi~is mildly supersonic. 
\item
Toward narrower channel width, the 1D power spectra of azimuthal profiles at the inner radii 
have a stronger shallowing trend than those at the outer radii, which implies 
that the shallower power spectra for narrower channel maps are mainly contributed by the inner 
disks, and thus the inner, more actively star-forming regions have proportionally more 
cooler \hi~than the outer regions.
\item
The power spectra of IC 1613 and IC 10, which are the two nearest galaxies in our sample, can be 
well fitted with a single power law between linear scales of $\sim$ 50 pc and $\sim$ 2 kpc.\ 
This suggests that the \hi~line-of-sight depth may be comparable with the maximum transverse 
scales in these two galaxies.\
\item
Our sample galaxies exhibit a bimodality in the spectral indices versus $M_{B}$ (and also SFR) plane.\ 
The division of the bimodality is at $M_{B} \sim$ $-$14.5 mag and log(SFR ($M_\odot\;{\rm yr}^{-1}$)) $\sim$ $-$2.1.\ 
Galaxies with higher luminosity and SFR tend to have shallower power spectra with a smaller scatter, 
whereas galaxies with lower luminosity and SFR exhibit a much larger scatter toward more negative 
spectral indices.\ The bimodality may signify that higher-luminosity galaxies tend to have bigger 
gas disks compared to their thicknesses, whereas lower-luminosity galaxies may be better described as 
thick disks or even triaxial ellipsoids.
\item
The inertial range spectral indices of single channel maps and velocity-integrated maps 
are not correlated with the SFR surface density.\ This may imply that either stellar and 
non-stellar energy sources can excite turbulence with about the same power spectral index, or 
non-stellar energy sources are more important in driving ISM turbulence.\ 
The single power-law behavior of the power spectra indicates that the ISM turbulence may be driven, 
from whatever energy sources (stellar or non-stellar), over a wide range of physical scales, 
otherwise we should see features, such as a flattening of power spectra at low wavenumbers, at the 
primary driving scale.\

\end{enumerate}
 
The multi-phase (i.e.\ different temperature components) nature of galactic neutral \hi~means that 
the power spectra determined for different velocity slice widths trace different temperature components 
of \hi.\ Therefore, determining the turbulent velocity spectral indices may be difficult, 
and more theoretical work taking into account the multi-phase nature of neutral \hi~is needed.\ 

\begin{acknowledgements}
This work was funded in part by the National Science Foundation through
grants AST-0707563 and AST-0707426 to DAH and BGE, and with general 
support from the National Radio Astronomy Observatory.\ HZ was partly supported 
by National Science Foundation of China through grants \#10833006 and \#11173059 to Yu Gao.\
We are grateful to the referee Frederic Bournaud for his very helpful suggestions, which 
significantly improved the paper.\ We thank Elias Brinks for his insightful suggestions and comments.\ 
We also thank Kimberly A. Hermann for providing helpful comments on the paper.\ 
\end{acknowledgements}

{\it Facilities: }\facility{NRAO}, \facility{{\it GALEX}}

                                                                                  
%
\begin{deluxetable}{lcccccccr} 
\tabletypesize{\scriptsize}                                                
\tablenum{1}                                                               
\tablecolumns{9}                                                           
\tablewidth{0pt}                                                           
\tablecaption{Galaxy Sample}
\tablehead{                                                                                                                
\colhead{Galaxy}           
& \colhead{Other Names}  
& \colhead{{\it D}}  
& \colhead{{\it M$_{B}$}}
& \colhead{log(SFR)} 
& \colhead{log({\it M$_{\rm atomic~gas}$})}  
& \colhead{\rh}
& \colhead{$R_{D}^V$}
& \colhead{Incl.} \\
 \colhead{}
& \colhead{} 
& \colhead{(Mpc)}     
& \colhead{(mag)}
& \colhead{\rm ($M_{\odot}$ yr$^{-1}$)}
& \colhead{\rm ($M_{\odot}$)}
& \colhead{(kpc)}  
& \colhead{(kpc)} 
& \colhead{(degrees)} \\                                                                 
\colhead{(1)}                                                                             
& \colhead{(2)}                                                                             
& \colhead{(3)}                                                                             
& \colhead{(4)}                                                                             
& \colhead{(5)}                                                                             
& \colhead{(6)}                                                                             
& \colhead{(7)}   
& \colhead{(8)}
& \colhead{(9)}                                                                                                                                                         
}                                                                          
\startdata                                                                                                               
DDO 210 \dotfill & Aquarius Dwarf & 0.9 &  $-$10.38 &  $-$4.07 &  6.52   & 0.78      &   0.17 &  47\\
M81dwA \dotfill & KDG 052 & 3.6 &  $-$11.46 &  $-$3.14 & 7.26   &   2.04 &   0.26 &   0\\
CVnIdwA \dotfill  & UGCA 292 & 3.6 & $-$12.16 & $-$2.68 & 7.81    & 2.22   & 0.57 &   41\\
DDO 155 \dotfill  & UGC 8091, GR 8 & 2.2 & $-$12.25 & $-$2.71 & 7.13   &  1.35    &   0.15 &   48\\
DDO 187 \dotfill & UGC 9128 & 2.2 &  $-$12.38 & $-$3.14 & 7.37    & 1.23  & 0.18 &   32 \\
DDO 53 \dotfill & UGC 4459, VIIZw 238 & 3.6 &  $-$13.43 & $-$2.34 & 8.39   &    3.14 &   0.72  &   43\\
DDO 75 \dotfill  & UGCA 205, Sextans A & 1.3 &  $-$13.72 & $-$2.18 & 8.01  &   3.09 &   0.22 &   34\\
F564-V3 \dotfill   & LSBC D564-08 & 8.7  & $-$13.67 & $-$3.05 & 7.56  &   3.37 &   0.53 &   53\\
DDO 70 \dotfill  & UGC 5373, Sextans B & 1.3 &  $-$13.74 & $-$2.59 & 7.72  &  3.22 &  0.48 &  24\\
NGC 4163 \dotfill & UGC 7199 & 2.9 & $-$13.95 & $-$1.73 & 7.34 & 0.60 & 0.10 & 0 \\
IC 1613 \dotfill & UGC 668, DDO 8 & 0.7 &  $-$14.17 & $-$2.23 & 7.66  &   2.66  &  0.58 &   38\\
Haro 29 \dotfill & UGCA 281, Mrk 209, IZw 36 & 5.9 &  $-$14.39 & $-$1.87 & 8.01  &  4.52  & 0.29 &  42\\ 
DDO 46 \dotfill & UGC 3966 & 6.1 & $-$14.36 & $-$2.05 & 8.35   &   4.76 &   1.14 &    29 \\
DDO 133 \dotfill & UGC 7698 & 3.5 & $-$14.36 & $-$2.13 & 8.23  &    3.82   &  1.14 &   43 \\
DDO 63 \dotfill  & UGC 5139, Holmberg I & 3.9 &  $-$14.58 & $-$2.09 & 8.33  &    4.22 &   3.09 &   0 \\
DDO 87 \dotfill & UGC 5918, KDG 072, VIIZw 347 & 7.7 &  $-$14.52 & $-$2.15 & 8.49  &    8.51 &   1.43 &    28 \\
DDO 101 \dotfill & UGC 6900 & 6.4 &  $-$14.40 & $-$2.68 & 7.10  &   2.36 &  0.93 &   49\\
DDO 43 \dotfill & UGC 3860 & 7.8 & $-$14.75 & $-$2.03 & 8.40    &  5.82 &   0.61 &   46 \\
DDO 52 \dotfill & UGC 4426 & 10.3 &  $-$15.05 & $-$2.03 & 8.57  &   6.99 &  1.32 &   53\\
DDO 47 \dotfill & UGC 3974 & 5.2 & $-$15.17 & $-$1.83 & 8.73  &   10.97 &   1.36 &    19 \\
IC 10 \dotfill & UGC 192 & 0.7 & $-$15.75 & $-$1.73\tablenotemark{a} & 8.16    & 4.00 &    0.40 &   41 \\
DDO 50 \dotfill  & UGC 4305, Holmberg II & 3.4 &  $-$16.39 & $-$1.18 & 8.99  &   8.67  &   1.10 &    47\\
NGC 3738 \dotfill & UGC 6565, Arp 234 & 4.9 &  $-$16.70 & $-$1.45 & 8.32   &  5.42 &  0.78 &   46 \\
NGC 4214 \dotfill  & UGC 7278 & 3.0 &  $-$17.26 & $-$1.03 & 8.91  &  8.62 &  0.75 &   26\\

\enddata

\tablecomments{
(1) Galaxy name.\
(2) The other commonly used names in the literature.\
(3) Distance from D. A. Hunter et al. (2012, in preparation and references therein).\
(4) $B$-band absolute magnitude.\
(5) Logarithm of SFR derived from the FUV luminosity (Hunter et al.\ 2010; Zhang et al.\ 2012).\ 
(6) Logarithm of the atomic gas mass (1.34$\times$\mh) is collected from single-dish observations 
in the literature (see Hunter \& Elmegreen 2004 for the references).\ 
(7) The radius where the \hi~column density falls off to 10$^{19}$ cm$^{-2}$.\ 
(8) Disk scale length measured on the $V$-band images.\ 
(9) The inclination angles derived by fitting the iso-intensity contours on the velocity-integrated \hi~maps.
}                                                            
\tablenotetext{a}{No FUV observations, so the SFR is derived from H$\alpha$ luminosity}           
\end{deluxetable}                                                          

                                                                                  
%
\begin{deluxetable}{lcccccccccc} 
\tabletypesize{\scriptsize}                                                
\tablenum{2}                                                               
\tablecolumns{11}                                                           
\tablewidth{0pt}                                                           
\tablecaption{Power Spectra}
\tablehead{                                                                                                                
\colhead{Galaxy}           
& \colhead{Min$_{\tt scale}$}   
& \colhead{Max$_{\tt scale}$} 
& \colhead{$<\beta_{\tt 1.3}>$}
& \colhead{$\sigma_{\beta}$}
& \colhead{$<\beta_{\tt 2.6}>$}
& \colhead{$\sigma_{\beta}$}
& \colhead{$\beta_{\tt integrated}$}
& \colhead{$\sigma_{\beta}$} 
& \colhead{Ch$_{\tt change}$}
& \colhead{$\beta_{\tt velocity}$} \\
 \colhead{}
& \colhead{(kpc)} 
& \colhead{(kpc)}     
& \colhead{}
& \colhead{}
& \colhead{}
& \colhead{}
& \colhead{}
& \colhead{} 
& \colhead{(km s$^{-1}$)} 
& \colhead{} \\                                                                 
\colhead{(1)}                                                                             
& \colhead{(2)}                                                                             
& \colhead{(3)}                                                                             
& \colhead{(4)}                                                                             
& \colhead{(5)}                                                                             
& \colhead{(6)}                                                                             
& \colhead{(7)}   
& \colhead{(8)}
& \colhead{(9)}                    
& \colhead{(10)}
& \colhead{(11)}                                                                                                                                      
}                                                                          
\startdata                                                                                                               
DDO 210 \dotfill  &     0.11 &     0.68 &     $-$2.17 &    0.09 &     $-$2.47 &    0.11 &     $-$3.66 &    0.05 & 15 &  $-$4.05\\
M81DWA \dotfill    &   0.20 &      1.35 &     $-$1.33 &    0.04 &     $-$1.38 &    0.04 &     $-$1.98 &    0.04 & 27  & $-$4.19\\
CVnIdwA \dotfill      & 0.38 &      2.37 &     $-$3.09 &    0.13 &     $-$3.47 &     0.14 &     $-$4.33 &    0.09 &  23 & $-$2.07\\
DDO 155 \dotfill     &  0.27 &      1.65 &     $-$2.59 &    0.12 &     $-$2.86 &     0.16 &     $-$3.97 &     0.15 & 6 & $-$3.28\\
DDO 187 \dotfill     &  0.11 &      0.43 &     $-$1.43 &    0.05 &     $-$1.48 &    0.06 &     $-$1.98 &    0.04 & 9 & $-$4.00 \\
DDO 53 \dotfill     &  0.23 &      2.23 &   \nodata  &  \nodata  &  $-$1.71 &    0.05 &       $-$2.48 &    0.04 & 15 & $-$4.54 \\
DDO 75 \dotfill     & 0.09 &      1.67 &    \nodata  &  \nodata  &  $-$2.15 &    0.04   &     $-$2.80 &    0.03 & 23 & $-$4.31 \\
F564-V3 \dotfill    &   1.31 &      5.09 &     $-$2.30 &    0.08 &     $-$2.46 &    0.08 &     $-$2.74 &     0.07 & 9 & $-$3.56 \\
DDO 70 \dotfill     &   0.14 &      1.61 &     $-$2.97 &    0.06 &     $-$3.29 &    0.08 &     $-$3.94 &     0.07 & 12 & $-$2.42 \\
NGC 4163 \dotfill    &   0.16 &      2.54 &     $-$1.73 &    0.04 &     $-$2.06 &    0.05 &     $-$3.32 &    0.04 & 24 & $-$4.88 \\
IC 1613 \dotfill   &   0.05 &      2.37 &     \nodata  &  \nodata  & $-$2.15 &    0.07 &       $-$2.85 &    0.02 & 26 & $-$4.38 \\
Haro 29 \dotfill    &   0.39 &      2.07 &     $-$1.80 &    0.05 &     $-$2.06 &    0.06 &     $-$2.87 &    0.07 & 17 & $-$4.62 \\
DDO 46 \dotfill    &   0.32 &      2.29 &     $-$1.59 &    0.04 &     $-$1.85 &    0.04 &     $-$2.75 &    0.02 & 14 & $-$4.80 \\
DDO 133 \dotfill    &   0.43 &      4.61 &     \nodata  &  \nodata  & $-$3.34 &    0.07 &      $-$3.94 &    0.10 & 13 & $-$2.31 \\
DDO 63 \dotfill   &    0.22 &      2.05 &    \nodata  &  \nodata  &  $-$1.44 &    0.04 &      $-$1.90 &    0.03 & 28 & $-$3.93 \\
DDO 87 \dotfill    &   0.48 &      4.50 &    \nodata  &  \nodata  &  $-$2.17 &    0.05 &       $-$2.90 &    0.03 & 15 & $-$4.46 \\
DDO 101 \dotfill   &    0.59 &      3.06 &    \nodata  &  \nodata  &  $-$2.05 &    0.08 &       $-$2.00 &     0.10 & 10 & $-$2.88 \\
DDO 43 \dotfill    &    0.65 &      5.87 &     \nodata  &  \nodata  & $-$2.13 &    0.06 &       $-$2.56 &    0.06 & 13 & $-$3.86 \\
DDO 52 \dotfill     &   0.85 &      4.93 &     $-$1.59 &    0.05 &     $-$1.93 &    0.05 &     $-$2.54 &    0.05 & 14 & $-$4.23 \\
DDO 47 \dotfill   &    0.42 &      3.91 &     \nodata  &  \nodata  & $-$2.30 &    0.05 &       $-$3.14 &    0.05 & 21 & $-$4.39 \\
IC 10 \dotfill       &     0.04 &       2.46 &     \nodata  &  \nodata  &  $-$1.45 &    0.04 &      $-$1.59 &    0.02 & 13 & $-$3.28 \\
DDO 50 \dotfill     &  0.25 &      2.72 &    \nodata  &  \nodata  &  $-$1.63 &    0.02 &      $-$2.25 &    0.02 & 26 & $-$4.25 \\
NGC 3738 \dotfill    &   0.32 &    3.04 &     $-$1.82 &    0.04 &     $-$2.13 &    0.05 &     $-$2.75 &    0.04 & 13 & $-$4.24 \\
NGC 4214 \dotfill    &   0.18 &      9.26 &     $-$1.92 &   0.03 &     $-$2.10 &   0.03 &     $-$2.31 &    0.05 & 14 & $-$3.43\\

\enddata

\tablecomments{
(1) Galaxy name.\ 
(2) The minimum linear scale in the power-law fitting to the power spectra.\ Min$_{\tt scale}$ is equal to 1.5 times the beam size.\ 
(3) The maximum linear scale in the power-law fitting to the power spectra.\
(4) The average 2D power spectral index of the 1.3 km s$^{-1}$ velocity slices.\
(5) The uncertainty of the power spectral index of the 1.3 km s$^{-1}$ velocity slices.\
(6) The average power spectral index of the 2.6 km s$^{-1}$ velocity slices.\
(7) The uncertainty of the power spectral index of the 2.6 km s$^{-1}$ velocity slices.\
(8) The power spectral index of the velocity-integrated map.\
(9) The uncertainty of the power spectral index of the velocity-integrated map.\
(10) The approximate channel width smaller than which the power spectra start to get shallower.\
(11) The velocity spectral index determined under the assumption that 1.3 km s$^{-1}$ velocity slices reach the thin regime 
and the line-of-sight depth is comparable to the maximum transverse scales.\ 
}                                                                       
\end{deluxetable}                                                          


\begin{thebibliography}{}

\bibitem[]{}Agertz, O., Lake, G., Teyssier, R., et al. 2009, MNRAS, 392, 294
\bibitem[]{}Banerjee, A., Jog, C. J., Brinks, E., \& Bagetakos, L. 2011, MNRAS, 415, 687
\bibitem[]{}Begum, A., Chengalur, J. N., \& Bhardwaj, S. 2006, MNRAS, 372, 33
\bibitem[]{}Bekki, K., \& Chiba, M. 2007, MNRAS, 381, 16
\bibitem[]{}Bekki, K., \& Stanimirovi\'c, S. 2009, MNRAS, 395, 342
\bibitem[]{}Bell, E. F., \& de Jong, R. S. 2001, ApJ, 550, 212
\bibitem[]{}Blitz, L., \& Rosolowsky, E. 2004, ApJ, 612, 29
\bibitem[]{}Boulares, A., \& Cox, D. P. 1990, ApJ, 365, 544
\bibitem[]{}Bournaud, F., Elmegreen, B. G., Teyssier, R., Block, D. L., \& Puerari, I. 2010, MNRAS, 409, 1088
\bibitem[]{}Braun, R. 1997, ApJ, 484, 637
\bibitem[]{}Burkhart, B., Stanimirovi\'c, S., Lazarian, A. \& Kowal, G. 2010, ApJ, 708, 1204
\bibitem[]{}Cannon, J. M., Most, H. P., Skillman, E. D., et al.\ 2011, ApJ, 735, 35
\bibitem[]{}Carignan, C., \& Purton, C. 1998, ApJ, 506, 125
\bibitem[]{}Chappell, D., \& Scalo, J. 2001, ApJ, 551, 712
\bibitem[]{}Cho, J., Lazarian, A., \& Vishniac, E. 2002, ApJ, 564, 291 
\bibitem[]{}Co\^t\'e, S., Carignan, C., \& Freeman, K. C. 2000, AJ, 120, 3027
\bibitem[]{}Combes, F., Boquien, M., Kramer, C., et al. 2012, A\&A, 539, 67
\bibitem[]{}Cornwell, T. J. 2008, IEEE Journal of Selected Topics in Signal Processing, 2, 793
\bibitem[]{}Crovisier, J., \& Dickey, J. M. 1983, A\&A, 122, 282
\bibitem[]{}de Blok, W. J. G., \& Walter, F. 2006, AJ, 131, 363
\bibitem[]{}Dib S., Bell, E., \& Burkert, A. 2006, ApJ, 638, 797
\bibitem[]{}Dickey, J. M., \& Brinks, E. 1993, ApJ, 405, 153
\bibitem[]{}Dickey, J. M., Mebold, U., Stanimirovi\'c, S., \& Staveley-Smith, L. 2000, ApJ, 536, 756
\bibitem[]{}Dutta, P., Begum, A., Bharadwaj, S., \& Chengalur, J. N. 2009a, MNRAS, 397, 60
\bibitem[]{}Dutta, P., Begum, A., Bharadwaj, S., \& Chengalur, J. N. 2009b, MNRAS, 398, 887
\bibitem[]{}Dickey, J. M., Mebold, U., Stanimirovi\'c, S., \& Staveley-Smith, L. 2000, ApJ, 536, 756
\bibitem[]{}Dickey, J. M., McClure-Griffiths, N. M., Stanimirovi\'c, S., Gaensler, B. M. \& Green, A. J. 2001, ApJ, 561, 264
\bibitem[]{}Efremov, Y. N., \& Elmegreen, B. G. 1998a, MNRAS, 299, 588
\bibitem[]{}Elmegreen, B. G., Kim, S., \& Staveley-Smith, L. 2001, ApJ, 548, 749
\bibitem[]{}Elmegreen, B. G., Elmegreen, D. M., \& Leitner, S. N. 2003, ApJ, 590, 271
\bibitem[]{}Elmegreen, B. G., \& Scalo, J. 2004, ARA\&A, 42, 211
\bibitem[]{}Federrath, C., Klessen, R. S., \& Schmidt, W. 2009, ApJ, 692, 364
\bibitem[]{}Field, G. B. 1965, ApJ, 142, 531
\bibitem[]{}Field, G. B., Goldsmith,D.W., \& Habing, H. J. 1969, ApJ, 155, 149
\bibitem[]{}Fournier, J. -D., \& Frisch, U. 1983, J. Mec. Theor. Appl., 2, 699
\bibitem[]{}Gazol, A., \& Kim, J. 2010, ApJ, 723, 482
\bibitem[]{}Gao, Y., \& Solomon, P. M. 2004, ApJ, 606, 271
\bibitem[]{}Geisler, D., Piatti, A. E., Bica, E., \& Clari\'a, J. J. 2003, MNRAS, 341, 771
\bibitem[]{}Gladwin, P. P., Kitsionas, S., Boffin, H. M. J., \& Whitworth, A. P. 1999, MNRAS, 302, 305
\bibitem[]{}Gibson, S. J., Taylor, A. R., Higgs, L. A., Brunt, C. M., \& Dewdney, P. E. 2005, ApJ, 626, 195
\bibitem[]{}Goldreich, P., \& Sridhar, S. 1995, ApJ, 438, 763
\bibitem[]{}Green, D. A. 1993, MNRAS, 262, 327
\bibitem[]{}Hennebelle, P., \& Audit, E. 2007, A\&A, 465, 431
\bibitem[]{}Heiles, C., \& Troland, T. H. 2003, ApJ, 586, 1067
\bibitem[]{}Hodge, P. W., \& Hitchcock, J. L. 1966, PASP, 78, 79
\bibitem[]{}Hopkins, P. F., Quataert, E., \& Murray, N. 2012, MNRAS, 421, 3488
\bibitem[]{}Hunter, D. A., \& Elmegreen, B. G. 2004, AJ, 128, 2170
\bibitem[]{}Hunter, D. A., \& Elmegreen, B. G. 2006, ApJS, 162, 49
\bibitem[]{}Hunter, D. A., Elmegreen, B. G., \& Ludka, B. C. 2010, AJ, 139, 447
\bibitem[]{}Jenkins, E. B., \& Tripp, T. M. 2001, ApJS, 137, 297
\bibitem[]{}Karachentsev, I. D., Karachentseva, V. E., Huchtmeier, W. K., \& Makarov, D. I. 2004, AJ, 127, 2031
\bibitem[]{}Khalil, A., Joncas, G., Nekka, F., Kestener, P., \& Arneodo, A. 2006, ApJS, 165, 512
\bibitem[]{}Kolmogorov, A. 1941, DoSSR, 30, 301
\bibitem[]{}Kowal, G., Lazarian, A., \& Beresnyak, A. 2007, ApJ, 658, 423
\bibitem[]{}Kim, J., \& Ryu, D. 2005, ApJ, 630, 45
\bibitem[]{}Kritsuk, A. G., \& Norman, M. L. 2002, ApJ, 569, L127
\bibitem[]{}Kritsuk, A. G., Norman, M. L., Padoan, P., \& Wagner, R. 2007, ApJ, 665, 416
\bibitem[]{}Krumholz, M. R., \& McKee, C. F. 2005, ApJ, 630, 250
\bibitem[]{}Lazarian, A., \& Pogosyan, D. 2000, ApJ, 537, 720 (LP00)
\bibitem[]{}Leroy, A. K., Walter, F., Brinks, E., et al. 2008, AJ, 136, 2782
\bibitem[]{}Lee, J. C., Kennicutt, R. C., Funes, S. J., et al. 2007, ApJ, 671, 113
\bibitem[]{}Lithwick, Y., \& Goldreich, P. 2001, ApJ, 562, 279
\bibitem[]{}Mac Low, M. M. 1999, ApJ, 524, 169
\bibitem[]{}Mac Low, M. M., \& Klessen, R. S. 2004, RvMP, 76, 125
\bibitem[]{}Muller, E., Staveley-Smith, L., Zealey, W. S., \& Stanimirovi\'c, S. 2003, MNRAS, 339, 105
\bibitem[]{}Nakamura, F., \& Li, Z. -Y. 2007, ApJ, 662, 395
\bibitem[]{}Norman, C. A., \& Ferrara, A. 1996, ApJ, 467, 280
\bibitem[]{}Ott, J., Walter, F., Brinks, E., et al. 2001, AJ, 122, 3070
\bibitem[]{}Padoan, P., \& Nordlund, A. 2002, ApJ, 576, 870
\bibitem[]{}Padoan, P., Juvela, M., Kritsuk, A., \& Norman, M. L. 2009, ApJL, 707, 153
\bibitem[]{}Puche, D., Westpfahl, D., Brinks, E., \& Roy, J.-R. 1992, AJ, 103, 1841
\bibitem[]{}Rich, J. W., de Blok, W. J. G., Cornwell, T. J., et al. 2008, AJ, 136, 2879
\bibitem[]{}Roychowdhury, S., Chengalur, J. N., Begum, A., \& Karachentsev, I. D. 2010, MNRAS, 404, 60
\bibitem[]{}Sargent, W. L. W., Sancisi, R., \& Lo, K. Y. 1983, ApJ, 265, 711
\bibitem[]{}S\'anchez-Janssen, R., M\'endez-Abreu, J., \& Aguerri, J. A. L. 2010, MNRAS, 406, 65
\bibitem[]{}Sellwood, J. A., \& Balbus, S. A. 1999, ApJ, 511, 660
\bibitem[]{}Simpson, C. E., Hunter, D. A., \& Knezek, P. M. 2005, AJ, 129, 160
\bibitem[]{}Spitzer, L. Jr. 1978, JRASC, 72, 349
\bibitem[]{}Stanimirovi\'c, S., Staveley-Smith, L., Dickey, J. M., Sault, R. J., \& Snowden, S. L. 1999, MNRAS, 302, 417
\bibitem[]{}Stanimirovi\'c, S., \& Lazarian, A. 2001, ApJ, 551, 53
\bibitem[]{}Staveley-Smith, L., Wilson, W. E., Bird, T. S., et al. 1996, PASA, 13, 243
\bibitem[]{}Sellwood, J. A., \& Balbus, S. A. 1999, ApJ, 511, 660
\bibitem[]{}Stone, J. M., Ostriker, E. C. \& Gammie, C. F. 1998, ApJ, 508, 99 
\bibitem[]{}Swaters, R. A., van Albada, T. S., van der Hulst, J. M., \& Sancisi, R. 2002, A\&A, 390, 829
\bibitem[]{}Tamburro, D., Rix, H.-W., Leroy, A. K., et al.\ 2009, AJ, 137, 4424
\bibitem[]{}van den Bergh, S. 1988, PASP, 100, 344
\bibitem[]{}Walter, F., \& Brinks, E. 1999, AJ, 118, 273
\bibitem[]{}Warren, S. R., Weisz, D. R., Skillman, E. D., et al. 2011, ApJ, 738, 1
\bibitem[]{}Wada, K., Meurer, G., \& Norman, C. A. 2002, ApJ, 577, 197
\bibitem[]{}Wolfire, M. G., McKee, C. F., Hollenbach, D., \& Tielens, A. G. G. M. 2003, ApJ, 587, 278
\bibitem[]{}Wong, T., \& Blitz, L. 2002, ApJ, 569, 157
\bibitem[]{}Young, L. M., \& Lo, K. Y. 1997, ApJ, 490, 710
\bibitem[]{}Zhang, Q., Fall, S. M., \& Whitmore, B. C. 2001, ApJ, 561, 727
\bibitem[]{}Zhang, H. -X., Hunter, D. A., Elmegreen, B. G., Gao, Y. \& Schruba, A. 2012, AJ, 143, 47


\end{thebibliography}
\end{document}